\documentclass[preprint,amsmath,amssymb,superscriptaddress,floatfix]{revtex4-2}
\usepackage{graphicx}
\usepackage{color}
\usepackage{hyperref}
\usepackage{makecell}
\usepackage{multirow}
\usepackage{booktabs}
\usepackage{longtable}

\begin{document}

\title{Finite-temperature properties and the hidden ferroelectric $R3c$ phase of bulk CaTiO$_3$ from second principles}
\author{Huazhang Zhang}
\email[Corresponding author: ]{zhanghz782@126.com}
\address{School of Physics and Mechanics, Wuhan University of Technology, Wuhan 430070, People's Republic of China}
\address{Theoretical Materials Physics, Q-MAT, University of Liège, B-4000 Sart-Tilman, Belgium}
\author{Michael Marcus Schmitt}
\address{Theoretical Materials Physics, Q-MAT, University of Liège, B-4000 Sart-Tilman, Belgium}
\author{Louis Bastogne}
\address{Theoretical Materials Physics, Q-MAT, University of Liège, B-4000 Sart-Tilman, Belgium}
\author{Xu He}
\address{Theoretical Materials Physics, Q-MAT, University of Liège, B-4000 Sart-Tilman, Belgium}
\author{Philippe Ghosez}
\email[Corresponding author: ]{Philippe.Ghosez@uliege.be}
\address{Theoretical Materials Physics, Q-MAT, University of Liège, B-4000 Sart-Tilman, Belgium}
\date{\today}

\begin{abstract}
A second-principles effective interatomic potential is introduced for the prototypical perovskite $\rm CaTiO_3$ (CTO), relying on a Taylor polynomial expansion of the Born-Oppenheimer energy surface around the cubic reference structure, in terms of atomic displacements and macroscopic strains. 
This model captures various phases of bulk CTO and successfully reproduces, in particular, the structure, energy, and dynamical properties of the nonpolar $Pbnm$ ground state as well as of the hidden ferroelectric $R3c$ phase. 
Finite-temperature simulations suggest that the still debated sequence of structural phase transitions over heating is $Pbnm \ (a^-a^-c^+) \rightarrow C2/m \ (a^-b^-c^0) \rightarrow I4/mcm \ (a^-c^0c^0) \rightarrow Pm\bar{3}m \ (a^0a^0a^0)$, a sequence during which the oxygen-octahedra rotations around the three pseudocubic axes vanish successively. 
Although never experimentally observed in bulk, the ferroelectric $R3c$ phase appears to be metastable and at an energy only slightly above the $Pbnm$ ground state at 0 K.
The simulations confirm that, if induced in some way, the $R3c$ phase remains stable up to about 300 K and shows ferroelectric properties. 
Furthermore, we find that the minimum energy path connecting the $Pbnm$ and $R3c$ phases involves localized layer-by-layer flipping of octahedral rotations, a mechanism which is shown to be at play during the thermal destabilization process of the $R3c$ phase toward the $Pbnm$ ground state.
The proximity of the $R3c$ phase with the $Pbnm$ ground state suggests that the former could be stabilized under electric field. 
However, due to the large energy barrier, the field required for the $Pbnm$-to-$R3c$ transition appears to be extremely large, consistent with the fact that bulk CTO was never reported to be ferroelectric nor antiferroelectric.
\end{abstract}

\maketitle

\newpage

\section{Introduction}
In the past several decades, solid-state research has been revolutionized by the advent of first-principles methods, particularly density functional theory (DFT) calculations. 
However, a major limitation of first-principles DFT calculations is their expensive electronic self-consistent iterations, which makes it challenging to apply this method to finite-temperature simulations of relatively large systems (more than hundreds of atoms).
The way to overcome this limitation is to integrate out the electronic degrees of freedom and to develop effective models that reproduce the potential energy surface (PES) as sampled from first-principles, from the so-called ``second-principles'' methods \cite{RN23}.
Generally, approaches for representing the PES can be roughly divided into two categories: one focuses on the local environments of atoms (e.g., shell models \cite{RN1069,RN1070,RN31,RN962}, bond-valence models \cite{RN13,RN14,RN3,RN19}, and various machine learning interatomic potentials \cite{RN992,RN165}), while the other considers global structural distortions with respect to a reference structure. 
The effective Hamiltonian method relying on lattice distortion modes is a pioneering example of the latter \cite{RN60,RN1072,RN70,RN58,RN1074,RN1067}. 
This method is particularly suitable for and has proven very useful in the study of ferroelectric compounds, as it captures the physical essence of the spontaneous polarization that appears from a small cooperative atomic distortion of the high-symmetry parent phase, as seen in displacive ferroelectrics such as BaTiO$_3$ \cite{RN60} and PbTiO$_3$ \cite{RN70}.
In the same direction, efforts have been made to represent the PES using a Taylor polynomial expansion in terms of individual real-space atomic displacements and macroscopic strains relative to a reference structure \cite{RN20,RN22,RN27,RN26,RN32,RN826,Yu2025,Louis2025}.
This more recent method naturally incorporates the complete set of lattice degrees of freedom and can be seen as a generalization of the effective Hamiltonian method, with potential applicability to more complex cases involving the couplings of multiple lattice distortions \cite{RN23}.

\begin{figure*}[htb]
\centering
\includegraphics[scale=0.50]{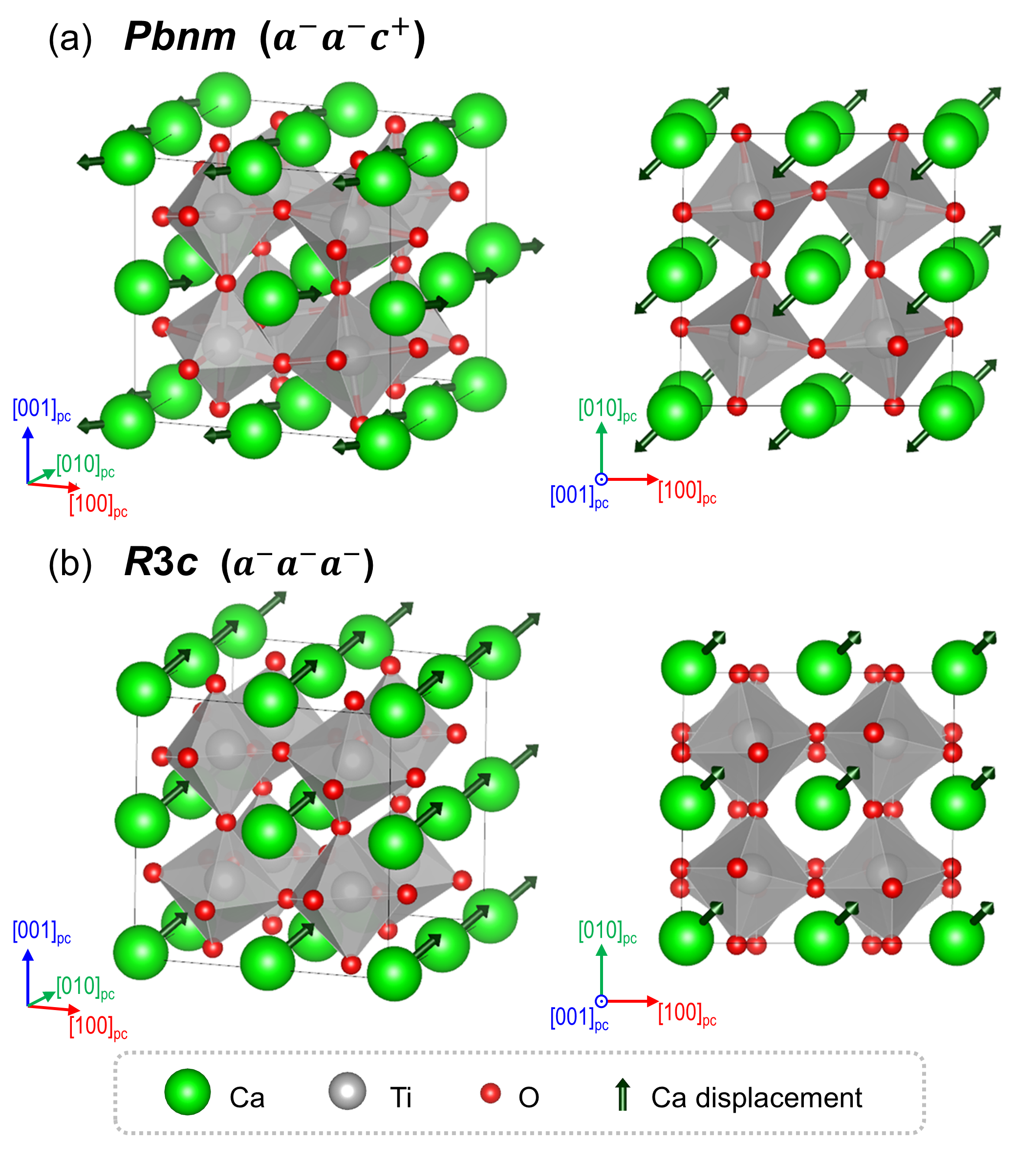}
\caption{Schematic views of the structures of the (a) $Pbnm$ and (b) $R3c$ phases of CTO.
The $Pbnm$ phase is featured by an octahedral rotation pattern $a^-a^-c^+$  and antipolar motions of the Ca atoms.
The $R3c$ phase has a different octahedral rotation pattern $a^-a^-a^-$ and polar displacements along the pseudocubic $[111]$ direction.}
\label{fig:structs}
\end{figure*}

CaTiO$_3$ (CTO) is the original mineral called ``perovskite'', which then gave its name to the related $AB$O$_3$ family of compounds.
CTO has a Goldschmidt tolerance factor $t < 1$, showing a strong tendency for octahedra rotations \cite{Goldschmidt1926}.
As most perovskites \cite{RN598}, CTO exhibits the nonpolar $Pbnm$ symmetry at low temperatures (below 1512 K, according to Ref. \cite{RN1018}).
The main feature of the $Pbnm$ phase is the $a^-a^-c^+$ oxygen octahedral rotations [Fig. \ref{fig:structs}(a)].
This phase also contains antipolar motions of Ca, which are induced by a trilinear coupling with the in-phase and antiphase octahedral rotations through the so-called ``hybrid-improper'' mechanism \cite{miao2013first}.
Upon heating, CTO undergoes a series of phase transitions before reaching the high-temperature $Pm\bar{3}m$ cubic phase (above 1636 K, according to Ref. \cite{RN1018}).
Although it is generally agreed that the tetragonal $I4/mcm$ phase precedes the cubic phase, the exact sequence of intermediate phases between $Pbnm$ and $I4/mcm$ remains debated.
Based on neutron diffraction and drop calorimetry measurements, Kennedy \textit{et al.} \cite{RN1055} and Guyot \textit{et al.} \cite{RN1057} suggested an intermediate phase with $Cmcm$ symmetry (rotation pattern $a^0b^-c^+$) between the $Pbnm$ and $I4/mcm$ phases. 
This has been reproduced by a Landau-type phenomenological potential fitted to the same experimental data \cite{RN487}. 
Independently, Carpenter suggested an intermediate $Imma$ phase (rotation pattern $a^-a^-c^0$) for the relevant (Ca, Sr)TiO$_3$ solid-solution system based on a theoretical Landau-type potential analysis \cite{RN1047}. 
Furthermore, temperature-dependent structural measurements by Yashima and Ali \cite{RN1018,RN1056} suggested that CTO may directly transform from the $Pbnm$ phase to $I4/mcm$ phase without an intermediate phase.
Overall, there is currently no consensus on the temperature-dependent phase transition sequence of CTO, so further investigations remain timely.

Although generally considered as a nonpolar material, CTO has recently attracted considerable research interest due to its hidden ferroelectricity. 
At very low temperatures, CTO exhibits characteristics of incipient ferroelectrics with the onset of a divergence in dielectric constant, suggesting the proximity of a ferroelectric transition \cite{RN502,RN1053}.
Additionally, specific ferroelastic domain walls in CTO have been found to be polar \cite{RN549,RN973,RN551,RN481}. 
Moreover, strain engineering has been shown to be capable of inducing global ferroelectricity in CTO \cite{RN467,RN483,RN487}.
Recently, Kim \textit{et al.} \cite{RN461} proposed an alternative strategy of octahedral rotation engineering, in which they used a LaAlO$_3$ $(111)$ substrate and successfully stabilized a strong ferroelectric $R3c$ phase with switchable polarization in ultrathin epitaxial CTO films at room temperature.
This $R3c$ phase is characterized by an octahedral rotation pattern $a^-a^-a^-$ and strong polar displacements along the pseudocubic $[111]$ direction [Fig. \ref{fig:structs}(b)], which give rise to a polarization of approximately 0.44 $\rm C/m^2$ \cite{RN461}.
Energetically, this $R3c$ phase is metastable: it is close to but higher in energy than the $Pbnm$ ground state, giving it a tendency to destabilize toward the $Pbnm$ ground state \cite{RN461,RN465,RN466}.
An interesting topic in this regard is the transition pathway between $R3c$ and $Pbnm$ phases, which, however, still needs to be fully clarified.

In this paper, we report the construction of a second-principles effective interatomic potential for CTO. 
The model was developed by reproducing first-principles data using MULTIBINIT which implements the scheme introduced by Wojde\l{} \textit{et al.} \cite{RN20}, in which the potential energy surface is represented through a Taylor polynomial expansion in terms of atomic displacements and macroscopic strains around a reference structure.
The model is thoroughly validated and is shown to correctly reproduce the relative stability and structural distortions of various phases, with particular success in capturing the nonpolar $Pbnm$ ground state and the strongly polar $R3c$ metastable phase.
Taking advantage of this model, the still-debated phase transition sequence of bulk CTO is reinvestigated. Then, the metastability of the polar $R3c$ phase is explored. The transition path between the metastable $R3c$ phase and the $Pbnm$ ground state is clarified and the possibility to induce the $R3c$ phase under electric field is investigated to assess the potential ferroelectric or antiferroelectric character of CTO in some temperature range.

\section{Methodology}
\subsection{First-principles calculations}
First-principles calculations were performed using the ABINIT package \cite{Abinit2025, Gonze2020, Gonze2016, Gonze2009, Gonze2002}.
We used the generalized gradient approximation (GGA) with Wu-Cohen (WC) parameterization for the exchange-correlation functional \cite{GGA_WC}, which has been shown to provide excellent results for CTO \cite{RN253}. 
We used optimized norm-conserving pseudopotentials \cite{NC_PSP} created with the WC functional treating the following orbitals as valence states: $3s$, $3p$, $4s$ for Ca, $3s$, $3p$, $3d$, $4s$ for Ti, and $2s$, $2p$ for O. 
The energy cutoff for the plane wave expansion was 40 hartrees, and $\Gamma$-centered Monkhorst-Pack \cite{monkhorst1976special} $k$-point mesh with a density of $8\times8\times8$ with respect to the five-atom cubic perovskite unit cell was used. 
For calculations to build the training set, $2\times2\times2$ supercells were used and the $k$-point mesh was reduced accordingly to $4\times4\times4$.
The dynamical matrices, Born effective charges, and dielectric tensor were calculated according to density functional perturbation theory (DFPT) as implemented in ABINIT \cite{DFPT}.

\subsection{Model construction}
The construction of the second-principles effective interatomic potential for CTO was carried out using MULTIBINIT \cite{Abinit2025, Gonze2020, MultibinitPaper, Multibinit}, which implements the second-principles approach outlined in Refs. \cite{RN20, RN22}.
The reference structure around which we develop the Taylor polynomial expansion is the cubic $Pm\bar{3}m$ phase. 
This structure is chosen because it has the highest symmetry among perovskite phases, which helps minimize the number of independent interaction terms.
Furthermore, the cubic phase is observed at high temperatures, and all other phases with lower symmetry can be derived from it.
The total interactions responsible for the energy gains with respect to the reference structure are split into the harmonic part (including long-range dipole-dipole and short-range harmonic interactions) and the anharmonic part, each of which was determined differently.
For the harmonic part, the model parameters, including interatomic force constants, Born effective charges, dielectric and elastic constants, were directly extracted from first-principles DFPT calculations on $8 \times 8 \times 8$ $q$-point mesh for the cubic reference structure.
For the anharmonic part, the following procedure was applied: 
First, a full set of symmetry-adapted terms (SATs) up to third- and fourth-order within an interatomic cutoff radius of one cubic lattice parameter (7.255 bohrs) was generated.
Second, the most relevant terms were selected to construct the model.
Third, the coefficients of the selected anharmonic SATs were determined by fitting the energy, forces, and stresses from a first-principles training set of atomic configurations constructed using $2\times2\times2$ supercells of the reference structure.
Finally, an automatic bounding algorithm was employed to include higher-order equivalent terms (sixth- and eighth-order) to ensure the boundedness of the effective potential \cite{MarcusPHD}.

\subsection{Model-based simulations}
Model-based simulations were performed with MULTIBINIT, which has been linked to various algorithms implemented in ABINIT.
Structural relaxations were performed using the Broyden-Fletcher-Goldfarb-Shanno (BFGS) algorithm with the convergence criteria of $10^{-4}$ $\rm hartrees/bohr$ on forces and $10^{-6}$ $\rm hartrees/bohr^3$ on stresses.
Finite-temperature simulations were performed using the hybrid Monte Carlo (HMC) approach \cite{HMC}, where the trial states for Metropolis Monte Carlo evaluation are generated through short molecular dynamics (MD) runs using NPT ensemble. 
We used 40 MD steps with a time step of 0.96 fs between each Monte Carlo evaluation.
At each temperature, the simulation was first run for no less than 1000 HMC sweeps to ensure that thermal equilibrium was reached, followed by additional 1000 HMC sweeps for averaging and property analysis.
Phonon dispersions were calculated according to the finite-difference method using ANADDB \cite{Anaddb} and ABIPY \cite{abipy}.
The minimum energy path between two stationary structures was calculated using the nudged elastic band (NEB) method \cite{NEB, SSNEB}, with minimization performed using the Quick-Min algorithm \cite{Quick_Min}.

\section{Model construction and validation}
\subsection{Model construction}

As explained in the methodology, the model consists of harmonic and anharmonic parts, and each part was constructed in different ways.
Accordingly, the model construction relies on two essential inputs from first-principles calculations:
(1) Derivative Database (DDB), a collection of second-order energy derivatives with respect to atomic displacements, strains, and electric field perturbations, providing the harmonic part of the model;
(2) Training Set (TS), a first-principles sampling of the lattice potential energy surface, used to fit and obtain the anharmonic coefficients of the model.

\begin{figure*}[htb]
\centering
\includegraphics[scale=0.48]{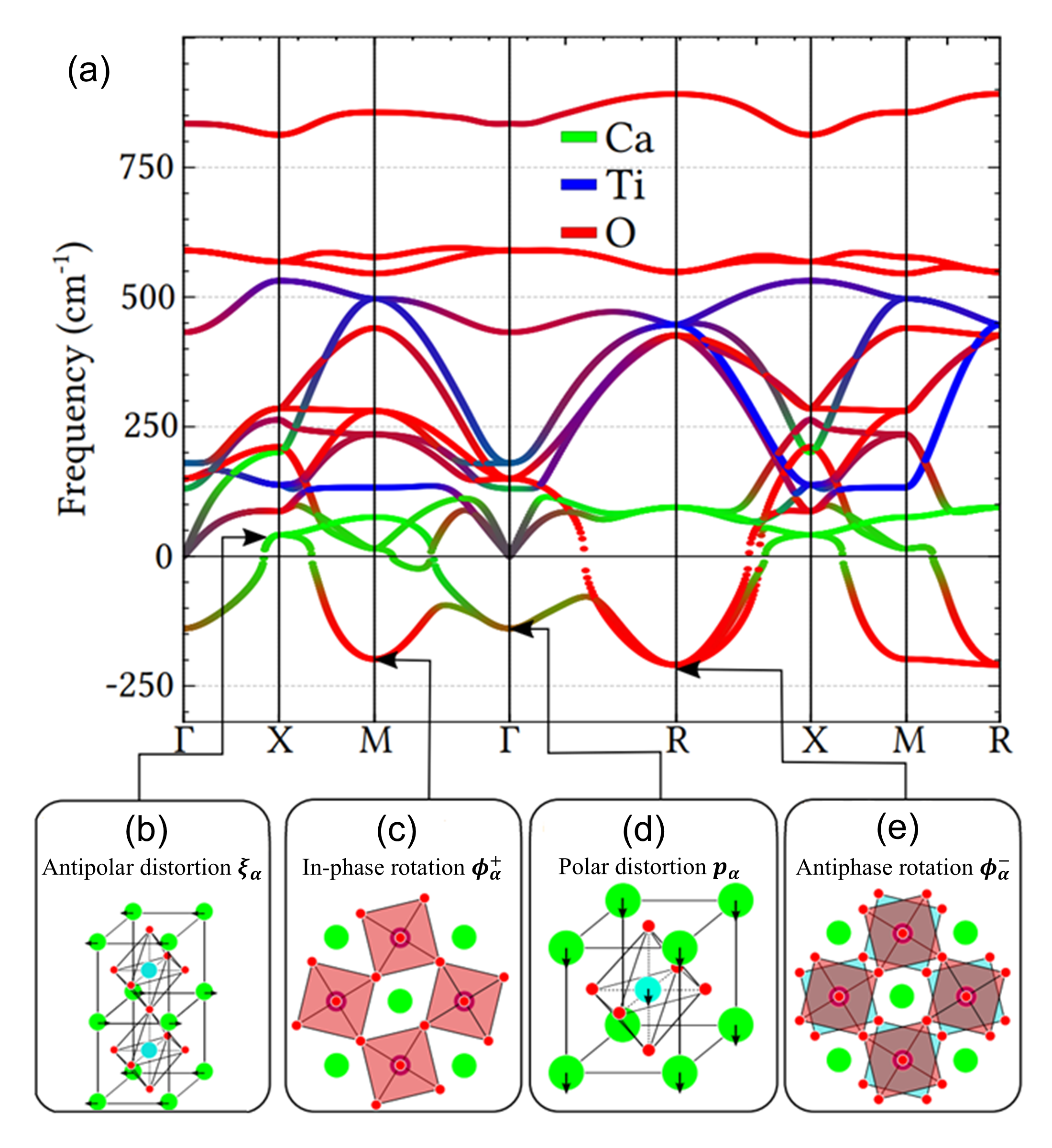}
\caption{(a) Phonon dispersions of the cubic reference structure of CTO. 
The colors represent the contributions of different atoms to the phonon eigendisplacements: green for Ca, blue for Ti, and red for O.
(b-e) Illustrations of the eigendisplacement patterns of important modes:
in-phase oxygen octahedral rotations $\phi_\alpha^+$ (the unstable mode at $M$),
antiphase oxygen octahedral rotations $\phi_\alpha^-$ (the unstable mode at $R$),
polar displacements $p_\alpha$ (the unstable mode at $\Gamma$), and
antipolar displacements $\xi_\alpha$ (the lowest-frequency mode at $X$),
where $\alpha$ denotes the Cartesian direction of the rotation axis or the polar or antipolar displacements.}
\label{fig:cubic_phonon}
\end{figure*}

\subsubsection{Harmonic part}
The harmonic part of the model, i.e. the DDB, was calculated directly from DFPT.
Fig. \ref{fig:cubic_phonon} shows the phonon dispersions of the cubic reference structure of CTO, which represent the second-order energy derivatives with respect to atomic displacements.
The strongest instabilities are the in-phase and antiphase rotations of the oxygen octahedra at the $M$ and $R$ points, respectively. 
In addition, although generally considered to be a nonpolar material, CTO also exhibits polar instabilities in its parent cubic phase, as indicated by the imaginary frequency at the $\Gamma$ point and consistent with a previous report \cite{RN253}.
These rotational and polar unstable modes of the cubic phase are also the prevalent lattice distortions observed in various distorted phases.
Moreover, there can be other lattice distortions appearing in distorted phases that are initially stable in the cubic phase. 
A typical example is the antipolar cation motions in the $Pbnm$ ground state of CTO, which correspond to $X$ point in the Brillouin zone of the cubic phase.
The condensation of the antipolar cation motions in the $Pbnm$ phase is due to the trilinear coupling with the in-phase and antiphase octahedral rotations.
Eventually, the comprehensive information regarding stable and unstable distortion modes in the cubic reference structure was incorporated into the model through its harmonic part.
Since the harmonic part was directly calculated by DFPT rather than being fitted, it remains exact with respect to the first-principles calculations.

\subsubsection{Anharmonic part}

\begin{figure*}[htb]
\centering
\includegraphics[scale=0.48]{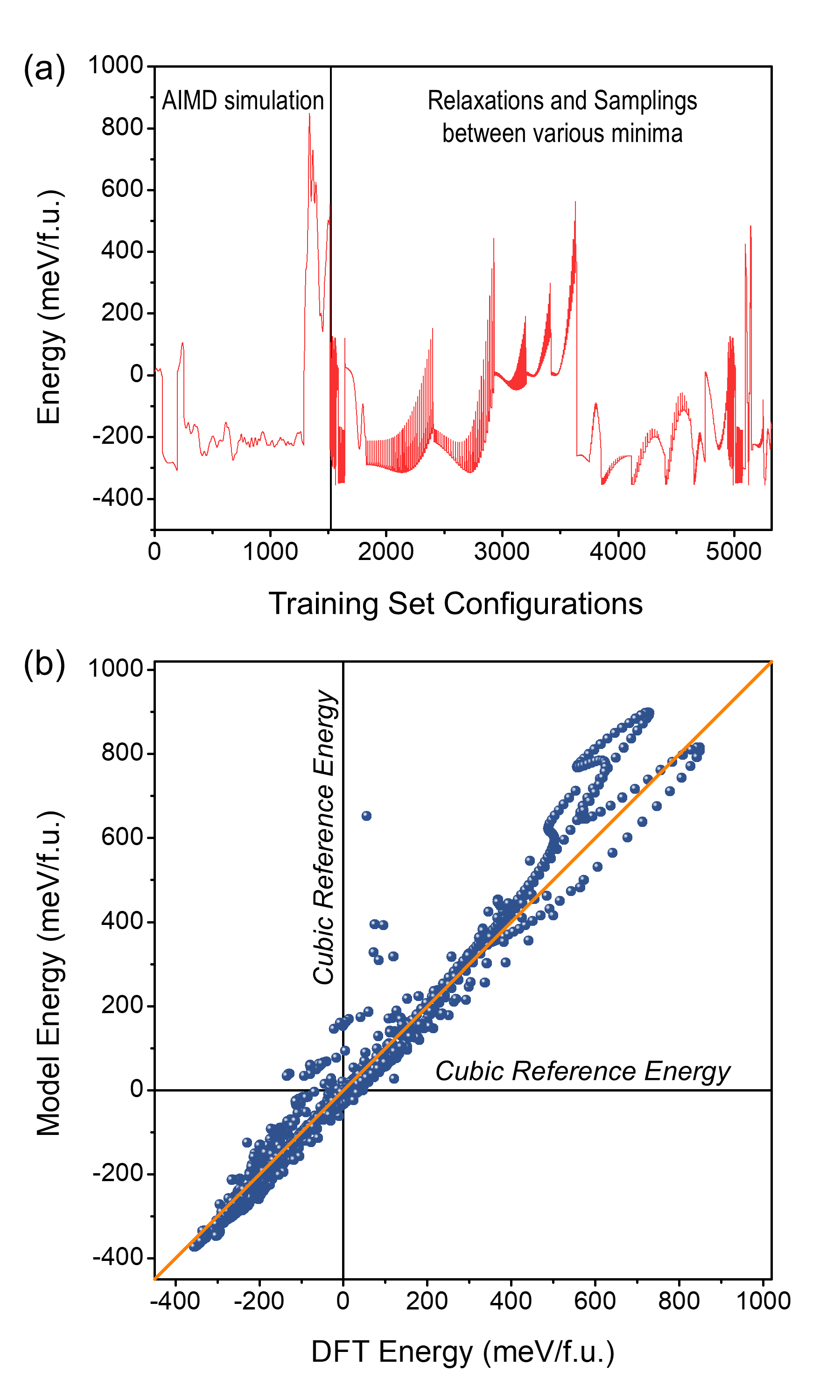}
\caption{(a) Energy profile of the configurations in the training set. 
(b) Reproduction of the training set configuration energies by the second-principles model.}
\label{fig:trainingset}
\end{figure*}

The anharmonic part of the model was obtained by fitting to a first-principles training set.
We used a hybrid strategy to build the training set for CTO.
Firstly, as in Ref. \cite{RN22}, we explored the dynamical distortions expected from first-principles by performing ab-initio molecular dynamics (AIMD) simulations. 
However, only this approach is insufficient for CTO, because the many important phases are almost never visited by the AIMD simulation, such as the phases with pure polar distortions $P4mm$, $Amm2$ and $R3m$. 
Thus, secondly, we sampled the potential energy surface by interpolations between the various stationary phases, and along these interpolation paths the lattice parameters were relaxed to get information about the strain-phonon couplings. 
Our final training set contains 5325 configurations. 
The energy profile of the training set configurations is shown in Fig. \ref{fig:trainingset}(a). 

To create, fit and select the anharmonic SATs, we used the MULTIBINIT software.
We find that a model containing 60 anharmonic SATs, with the first 30 terms having Ca as the common atom and the subsequent 30 terms having Ti as the common atom, is sufficient to well reproduce the training set and correctly capture the most important phases.
However, using this 60-term model in finite-temperature simulations frequently encounters divergence problems because the model energy was unbounded from below.
Therefore, we developed a bounding algorithm that introduces higher-order bounding terms, which have even orders in all atomic displacements and strains and have positive coefficient values, based on the terms contained in the unbounded model \cite{MarcusPHD}.
After the bounding procedure, 179 additional anharmonic terms were added into the model, and any possible divergence was effectively eliminated, without significantly deteriorating the quality of the fit.
Fig. \ref{fig:trainingset}(b) shows the reproduction of the training set by the bounded model.
Besides a few outliers, we find a fairly good linear relationship between the DFT energy and the model-predicted energy for the training set configurations, with $R^2 = 0.9661$, showing an acceptable reproducibility of the training set by the model.  
The final second-principle CTO model contains 239 anharmonic terms. 
All these SATs and their coefficient values are listed in the Appendix, Table \ref{tab:coeffs}.

\subsection{Model validation}

\subsubsection{Stationary phases}

\begin{table*}[b]
\small
\caption{
Comparisons of energy and lattice parameters of different phases of CTO between DFT and model-based structural relaxations.
The energies are reported in relative values with respect to the cubic reference structure.}
\label{tab:relax}
\begin{ruledtabular}
\begin{tabular}{ccccccccc}
 \multirow{2}{*}{Space group}  &
 \multicolumn{4}{c}{DFT}  &
 \multicolumn{4}{c}{Multibinit}  \\
\cmidrule(lr){2-5}
\cmidrule(lr){6-9}
 & $E$ (meV/f.u.) & $a$ (\AA) & $b$ (\AA) & $c$ (\AA) 
 & $E$ (meV/f.u.) & $a$ (\AA) & $b$ (\AA) & $c$ (\AA) \\
\colrule
 $R3m$         &  -42.0 & 5.422 & 5.422 &  6.751  &  -60.6 & 5.427 & 5.427 &  6.750  \\
 $Amm2$        &  -48.4 & 3.821 & 5.423 &  5.545  &  -74.6 & 3.815 & 5.441 &  5.541  \\
 $P4mm$        &  -72.1 & 3.803 & 3.803 &  4.018  &  -91.4 & 3.810 & 3.810 &  3.974  \\
 $P4/mbm$      & -224.2 & 5.338 & 5.338 &  3.882  & -223.3 & 5.362 & 5.362 &  3.865  \\
 $Pmc2_1$      & -238.0 & 3.842 & 5.383 &  5.410  & -240.6 & 3.845 & 5.401 &  5.402  \\
 $I4/mcm$      & -261.1 & 5.336 & 5.336 &  7.750  & -278.8 & 5.353 & 5.353 &  7.735  \\
 $R\bar{3}c$   & -280.3 & 5.405 & 5.405 & 13.122  & -315.9 & 5.381 & 5.381 & 13.213  \\
 $Cmcm$        & -291.6 & 7.564 & 7.643 &  7.635  & -315.1 & 7.556 & 7.665 &  7.634  \\
 $Imma$        & -296.2 & 5.401 & 7.566 &  5.407  & -318.7 & 5.415 & 7.546 &  5.405  \\
 $R3c$         & -303.3 & 5.395 & 5.395 & 13.313  & -349.0 & 5.386 & 5.386 & 13.367  \\
 $Pbnm$        & -355.3 & 5.433 & 7.587 &  5.333  & -375.3 & 5.429 & 7.600 &  5.346
\end{tabular}
\end{ruledtabular}
\end{table*}

\begin{figure*}[htb]
\centering
\includegraphics[scale=0.55]{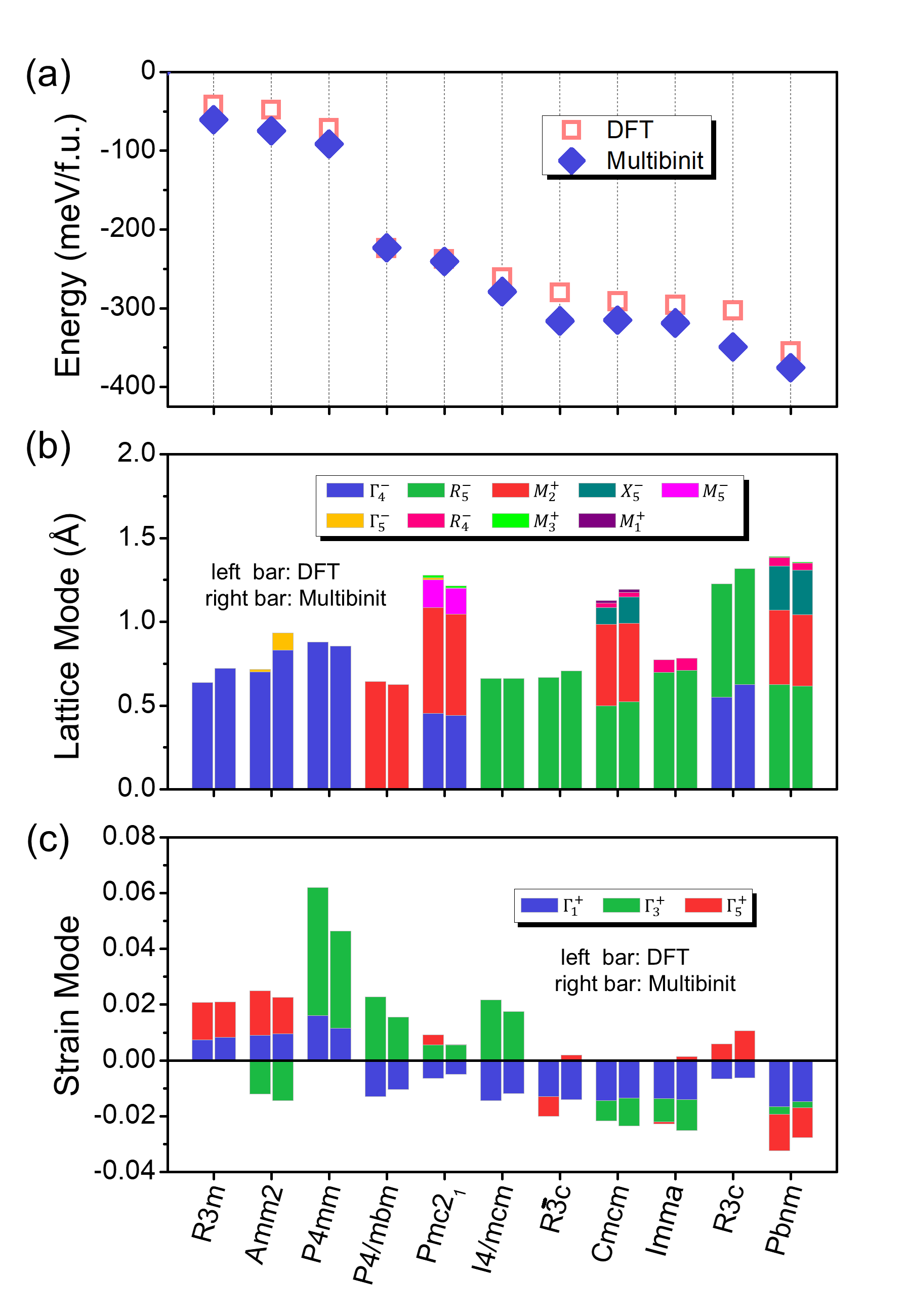}
\caption{Comparisons of (a) energy, (b) lattice distortion modes and (c) lattice strains of different phases of CTO between DFT and model-based structural relaxations.
The symmetry-adapted lattice distortion modes and strain modes were analyzed with ISODISTORT \cite{isodistort}.
The amplitudes of the lattice distortion modes are reported in the so-called ``parent-cell-normalized'' values.
For the strain modes, $\Gamma_1^+$, $\Gamma_3^+$, and $\Gamma_5^+$ correspond to volumetric strain, orthogonal nonvolumetric strain, and shear strain, respectively.}
\label{fig:relaxations}
\end{figure*}

To verify to which extent the model successfully captures the important metastable phases, we performed structural relaxations and compared the resulting energies and lattice distortions with those calculated by DFT relaxations. 
As illustrated in Fig. \ref{fig:relaxations}(a), while the model slightly overstabilizes the energy of some phases, it successfully reproduces the correct energetic order. 
Fig. \ref{fig:relaxations}(b) compares the symmetry-adapted lattice distortion modes condensed in different distorted phases between DFT and model-based structural relaxations.
It is observed that all the lattice distortion modes appearing in the DFT-relaxed structures also appear in the model-relaxed structures, with comparable mode amplitudes in both cases. 
Fig. \ref{fig:relaxations}(c) shows that the model also reproduces well the lattice strains, with few exceptions in the shear strain modes for some specific phases (e.g. $R\bar{3}c$ and $Imma$). 
In addition, Table \ref{tab:relax} shows that the lattice parameters of different phases calculated from DFT and the model exhibit fairly good agreement.
These results indicate that the key stationary points on the potential energy surface are well captured by the model.

\subsubsection{Phonon dispersions of distorted phases}

\begin{figure*}[b]
\centering
\includegraphics[scale=0.55]{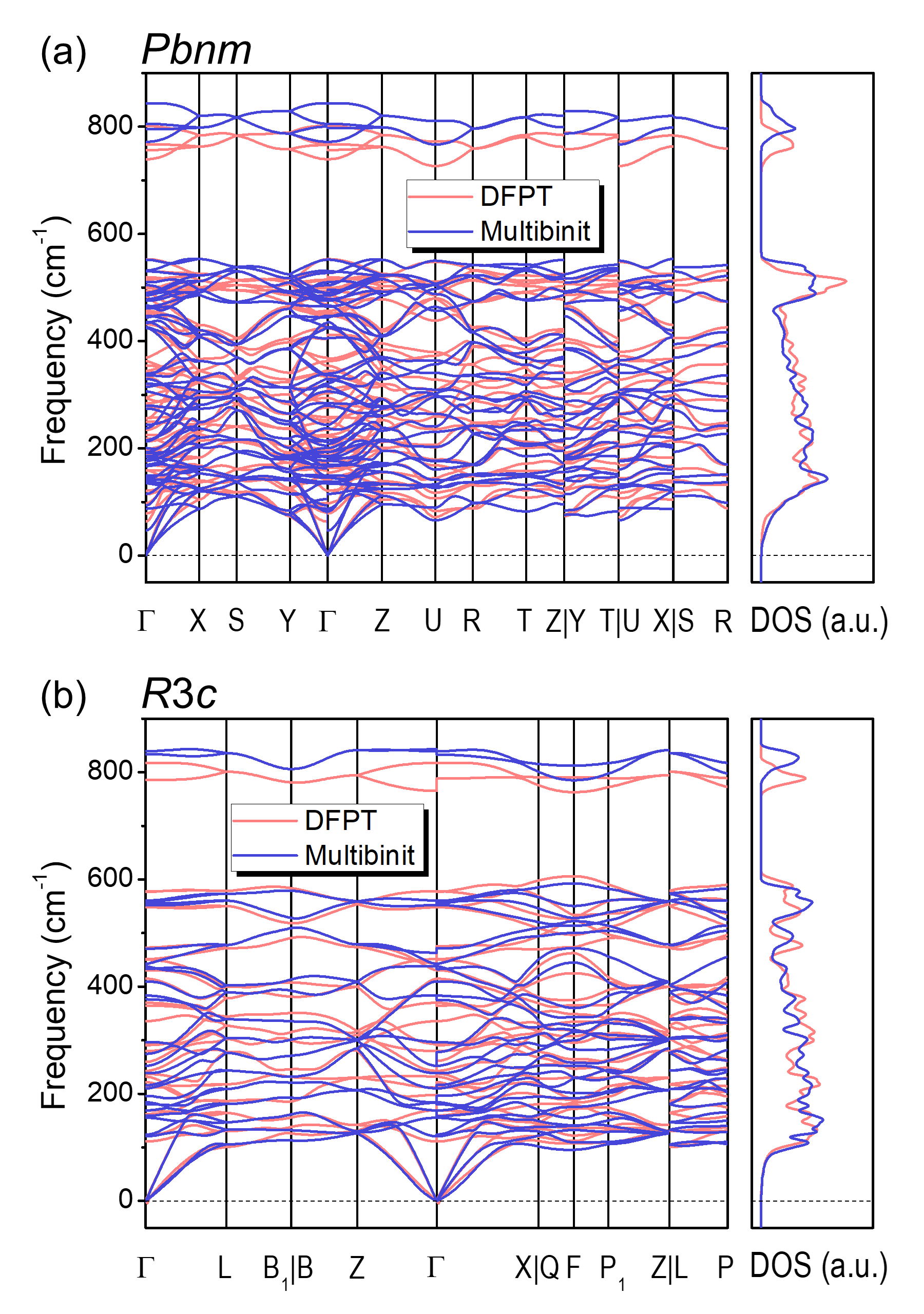}
\caption{Comparisons of phonon dispersions and phonon density of states between DFPT and the mode-based finite-difference calculations for (a) the $Pbnm$ ground state and (b) the $R3c$ metastable phase.}
\label{fig:phonon}
\end{figure*}

Benefiting from the inclusion of full lattice degrees of freedom, the model allows the computation of complete phonon dispersions.
Since the harmonic part of the model is accurate with respect to first-principles calculations, the phonon dispersions of the cubic reference structure are also exact.
A meaningful validation of the model is comparing the phonon dispersions and the phonon density of states of the distorted phases, with those obtained from DFPT.
Fig. \ref{fig:phonon} provides such comparisons around the $Pbnm$ ground state and the $R3c$ metastable phase.
Before calculating the phonon dispersions, the structures were fully relaxed using DFT and the model, respectively.
It can be seen that although the effective potential is developed using Taylor expansion around the cubic reference structure, rather than around the highly distorted $Pbnm$ and $R3c$ phases, the model still accurately predicts the phonon dispersions for these distorted phases.
These results demonstrate that the curvatures of the potential energy surface near these distorted phases are also well captured.

\section{Temperature-dependent phase transitions}
The previous model can now be used to investigate the temperature-dependent phase transitions of CTO.
We perform HMC simulations in $16 \times 16 \times 16$ supercell, starting from the low-temperature $Pbnm$ ground state and progressively heating up until the system becomes cubic.
Throughout the heating process, the system stays nonpolar. 
To monitor structural phase transitions, we track the evolution of octahedral rotations and lattice parameters as functions of temperature.

\begin{figure*}[t]
\centering
\includegraphics[scale=0.55]{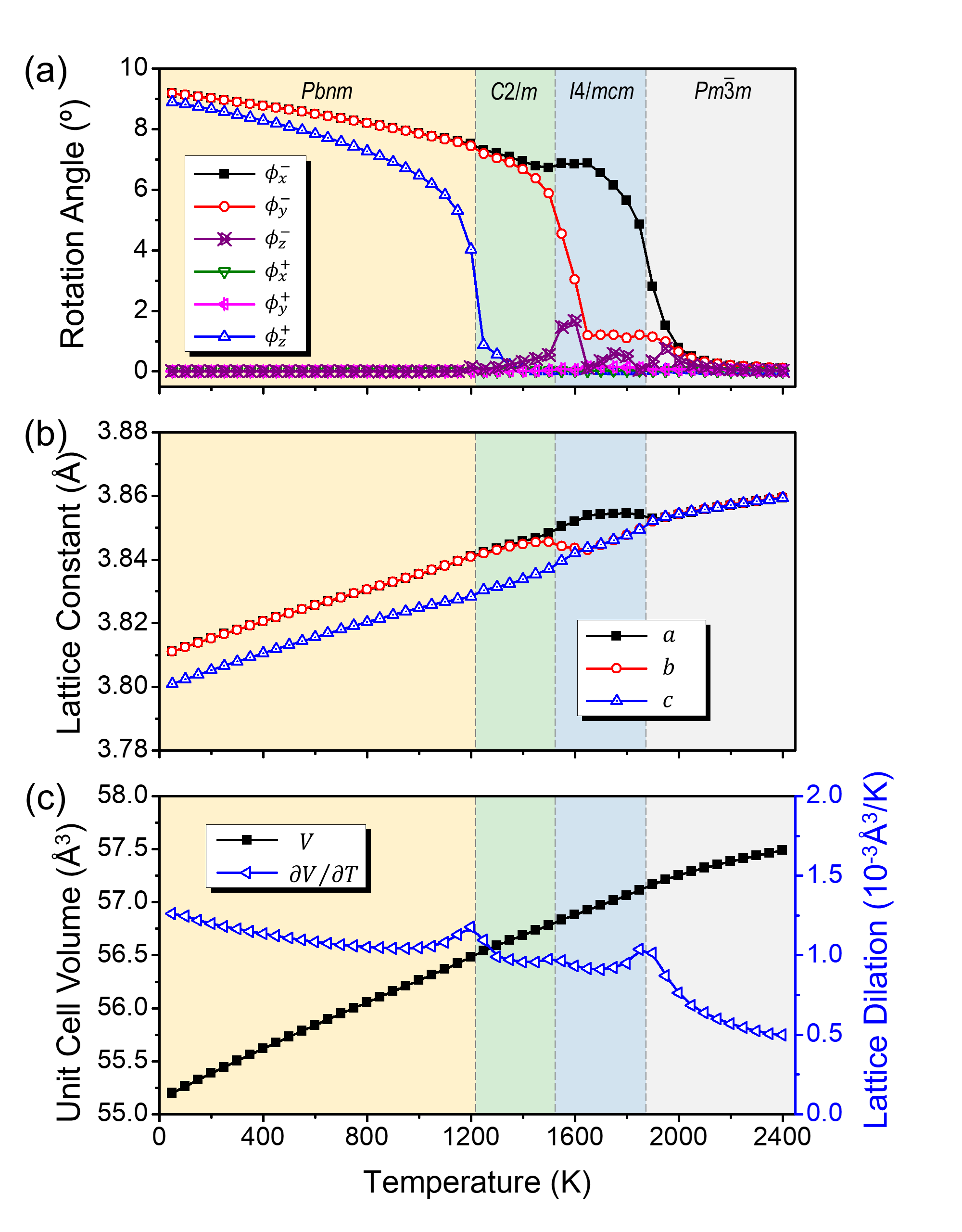}
\caption{Structural phase transitions during the heating process from the $Pbnm$ ground state. 
(a) Oxygen octahedral rotations, (b) pseudocubic lattice parameters, and (c) unit cell volume and lattice dilation as functions of temperature. 
The simulation is performed in a $16 \times 16 \times 16$ supercell using the second-principles model of CTO.}
\label{fig:Pnma_heating}
\end{figure*}

The model-based finite-temperature simulation shows that the system stays in the $Pbnm$ phase at low temperatures, characterized by the $a^-a^-c^+$ rotation pattern and the lattice parameters $a = b > c$ [Figs. \ref{fig:Pnma_heating}(a) and (b)].
With the increase of temperature, several phase transitions can be observed, as indicated by the changes in octahedral rotation pattern and lattice parameters.
In addition, the volume of the unit cell shows rapid increases near the phase transitions, resulting in peaks in lattice dilation curve [Fig. \ref{fig:Pnma_heating}(c)].
The phase transition temperatures are estimated by these peaks in the lattice dilation curve.

Upon heating from the low-temperature $Pbnm$ phase, the first phase transition occurs around 1200 K, where the in-phase rotation $\phi_z^+$ exhibits a sudden reduction and the antiphase rotations $\phi_x^-$ and $\phi_y^-$ are almost unchanged.
Above this temperature, the antiphase rotations $\phi_x^-$ and $\phi_y^-$ have slightly different amplitudes, suggesting that the phase above the first transition should be assigned to be $C2/m$ with $a^-b^-c^0$ rotation pattern.
This $C2/m$ phase is very similar to the $Imma$ ($a^-a^-c^0$) phase suggested by Carpenter \cite{RN1047}; the main difference lies in the amplitudes of the antiphase rotations about the two pseudocubic axes, which are equal in the $Imma$ phase but unequal in the $C2/m$ phase.

With the temperature further increasing, the next phase transition occurs around 1550 K, where $\phi_y^-$ rapidly decreases and $\phi_x^-$ slightly increases. 
Above this temperature, only $\phi_x^-$ has a significant amplitude, indicating that the phase is $I4/mcm$ with $a^-c^0c^0$ rotation pattern. 
Finally, around 1900 K, the amplitude of the last rotation $\phi_x^-$ decreases toward zero, indicating that the system changes to the cubic $Pm\bar{3}m$ phase.
Therefore, the finite-temperature simulation suggests a phase transition sequence for CTO: $Pbnm \ (a^-a^-c^+) \rightarrow C2/m \ (a^-b^-c^0) \rightarrow I4/mcm \ (a^-c^0c^0) \rightarrow Pm\bar{3}m \ (a^0a^0a^0)$, where the octahedral rotations about the three pseudocubic axes disappear progressively.

\begin{figure*}[bt]
\centering
\includegraphics[scale=0.50]{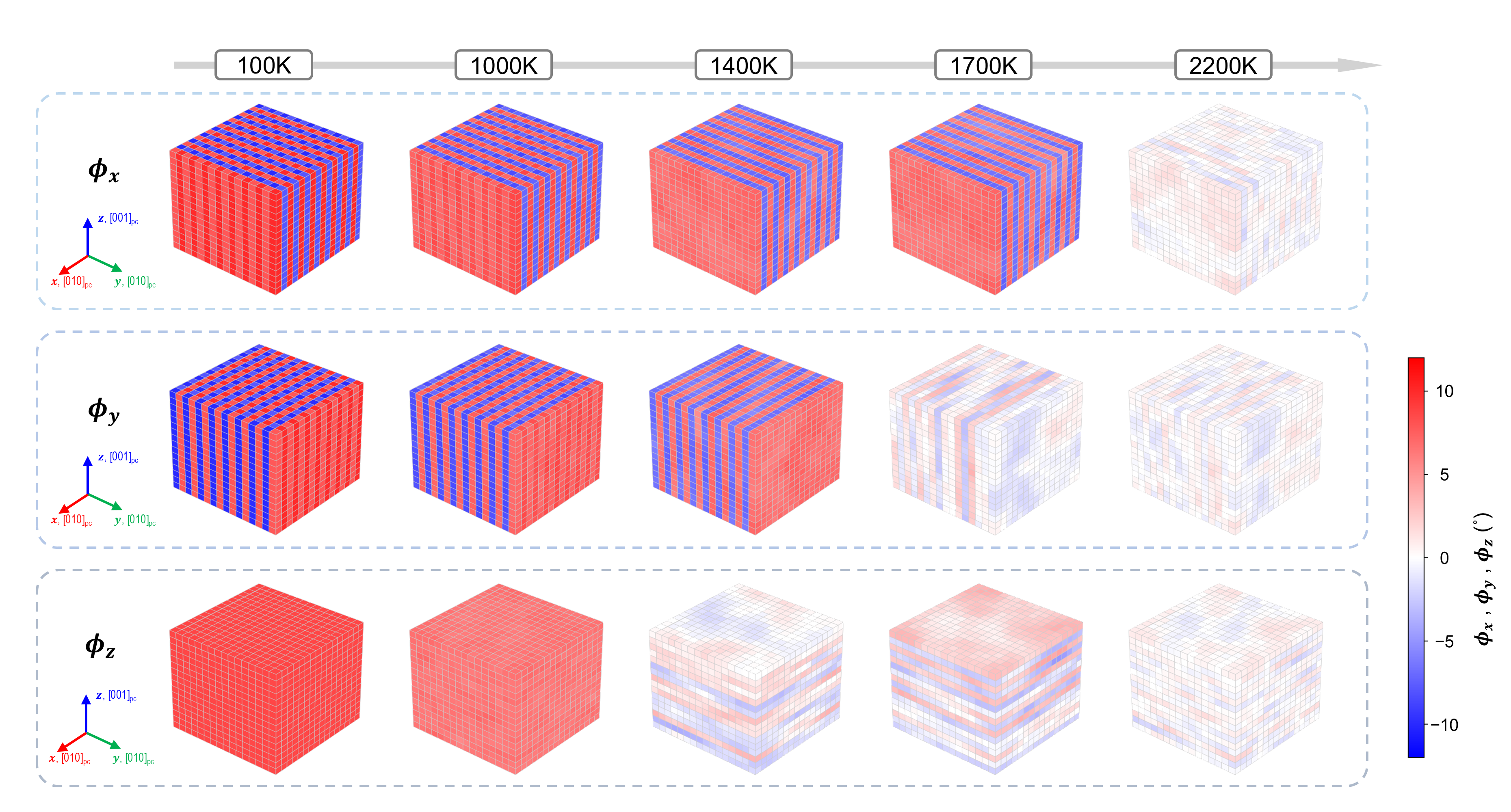}
\caption{Evolution of octahedral rotations during the heating process from the $Pbnm$ ground state.
The intensity of the colors represents the amplitude of $\phi_x$, $\phi_y$, and $\phi_z$ defined by Eq. (\ref{eq:phi}), with red and blue indicating opposite signs.}
\label{fig:Pnma_rot}
\end{figure*}

The evolution of octahedral rotations with temperature is further illustrated in Fig. \ref{fig:Pnma_rot}, where the three-dimensional distributions of local oxygen octahedral rotations in the $16 \times 16 \times 16$ simulating supercell are depicted. 
The rotations of the local oxygen octahedra are characterized using $\phi_x$, $\phi_y$, $\phi_z$ as defined by
\begin{equation}
\phi_x = \omega_x (-1)^{i_y+i_z}, \quad
\phi_y = \omega_y (-1)^{i_z+i_x}, \quad
\phi_z = \omega_z (-1)^{i_x+i_y},
\label{eq:phi}
\end{equation}
where $\omega_x$, $\omega_y$, $\omega_z$ is the rotation angle around the pesudocubic $[100]$, $[010]$, and $[001]$ directions, respectively, and $i_x$, $i_y$, $i_z$ are integers that indicate the position of the local oxygen octahedra in the directions of the three pseudocubic axes.
As shown in Fig. \ref{fig:Pnma_rot}, at low temperatures (e.g., 100 K), $\phi_x$ and $\phi_y$ alternate between positive and negative values, indicating that the rotations around the pesudocubic $[100]$ and $[010]$ axes are antiphase.
In contrast, $\phi_z$ does not change its sign throughout the supercell, indicating that the rotations around the pesudocubic $[001]$ axis are in-phase.
This is consistent with the $a^-a^-c^+$ rotation pattern in the low-temperature $Pbnm$ phase.
With the increase of temperature, the octahedral rotations around the three pseudocubic axes progressively diminish: first, the in-phase rotations around the pesudocubic $[001]$ axis disappear, followed by the disappearance of antiphase rotations around the pesudocubic $[010]$ axis, and then $[100]$ axis. 
Additionally, above those transition temperatures, local octahedral rotations with small amplitude persist and exhibit a layered-correlation feature. 
This layered-correlation feature of octahedral rotations is related to the geometric connectivity of the octahedra in the perovskite structure, which imparts strong correlation of octahedral rotations within the plane perpendicular to the rotation axis and weak correlation of the octahedral rotations between adjacent planes.

As mentioned in the Introduction, there is general consensus from experiments on the low-temperature $Pbnm$ phase, and the high-temperature $I4/mcm$ and $Pm\bar{3}m$ phases, while the potential intermediate phase between the $Pbnm$ and $I4/mcm$ phases is still under debate.
Our simulation is consistent with experiments with regard to the $Pbnm$, $I4/mcm$, and $Pm\bar{3}m$ phases, and predicts an intermediate $C2/m$ phase between the $Pbnm$ and $I4/mcm$ phases.
To compare with experimental phase transition temperatures, we note that the model predicts the disappearance of in-phase and antiphase octahedral rotations at approximately 1200 K [where the transition $Pbnm \ (a^-a^-c^+) \rightarrow C2/m \ (a^-b^-c^0)$ occurs] and 1900 K [where the transition $I4/mcm \ (a^-c^0c^0) \rightarrow Pm\bar{3}m \ (a^0a^0a^0)$ occurs], respectively. In comparison, x-ray and neutron diffraction experiments \cite{RN1055,RN1056,RN1018} show that superlattice reflections associated with in-phase octahedral rotations vanish around 1500 K, while those related to antiphase rotations disappear around 1600 K.
Therefore, our model appears to underestimate the temperature at which in-phase octahedral rotations vanish and overestimate the temperature at which antiphase rotations disappear.

\section{Metastable $R3c$ phase}
Although the polar $R3c$ phase has a higher energy than the $Pbnm$ ground state, it is free of lattice instabilities [Fig. \ref{fig:phonon}(b)], so that it corresponds to a local minimum on the potential energy surface and has the potential to be stabilized at low temperatures. 
Consistently, this polar $R3c$ phase has been experimentally stabilized in CTO films at room temperature \cite{RN461}.
Our second-principles CTO model successfully captures this metastable $R3c$ phase, providing an opportunity to further investigate its properties.
While our model confirms that, as observed experimentally \cite{RN461}, the $R3c$ phase can be stabilized  by epitaxial strain and octahedral rotation couplings in epitaxial CTO films on (111)-LaAlO$_3$ substrates \cite{Zhang_unpublished}, here we better focus on the intrinsic metastability of the $R3c$ phase in bulk CTO and explore how it can be eventually stabilized under external field, probing and assessing the potential ferroelectric or antiferroelectric nature of CTO.

\subsection{Minimum energy path connecting $Pbnm$ and $R3c$ phases}

\begin{figure*}[b]
\centering
\includegraphics[scale=0.40]{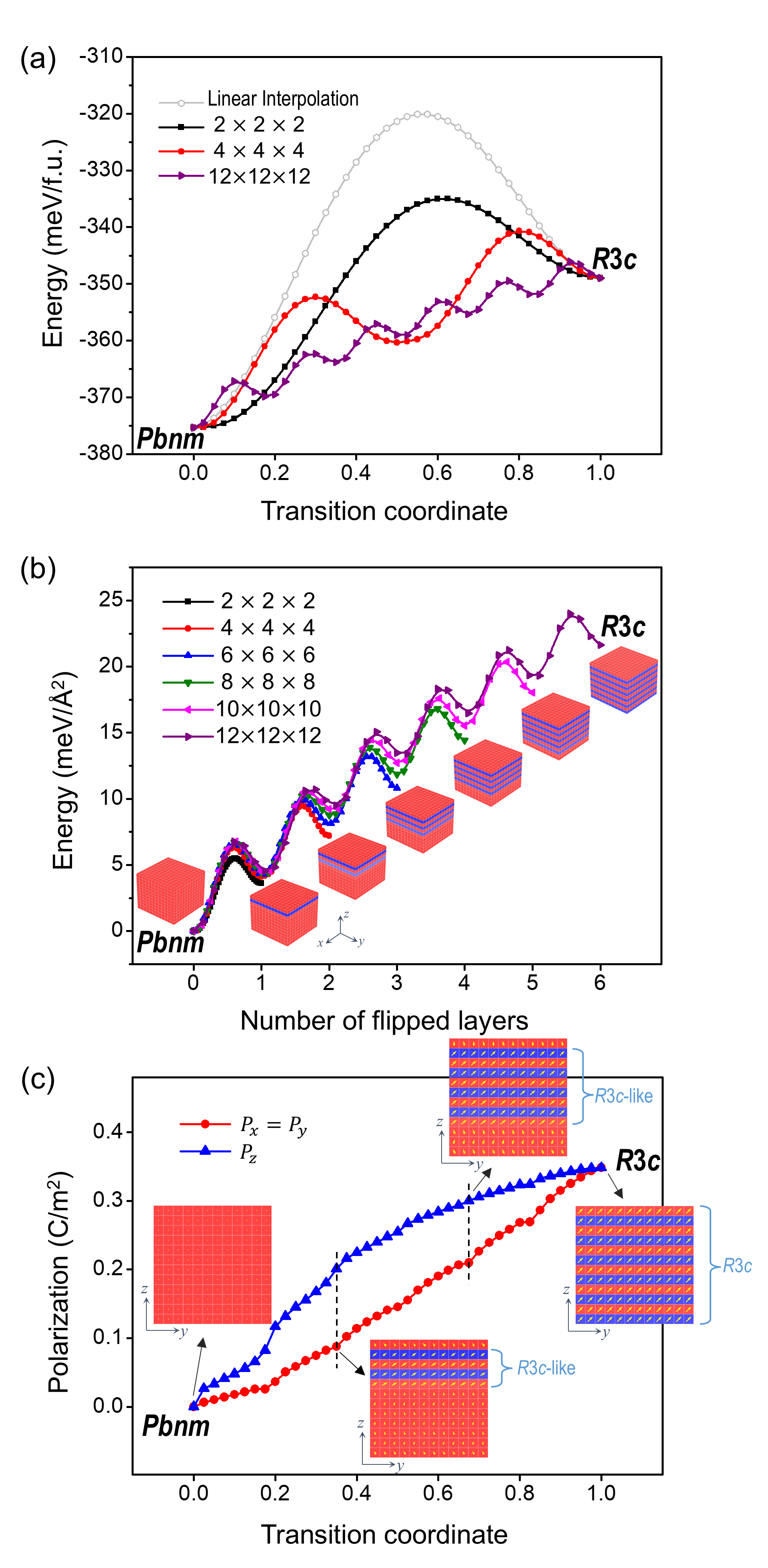}
\caption{Minimum energy path connecting the $Pbnm$ and $R3c$ phases calculated using the NEB method on supercells of different sizes.
(a) Energy normalized over the size of the entire simulation system (energy per formula unit) and plotted against the transition coordinate.
(b) Energy normalized by cross-sectional area and plotted against the number of flipped layers of octahedra rotations.
The inset cubes in (b) depict the evolution of octahedral rotations around the pseudocubic [001] axis, characterized by $\phi_z$ as defined by Eq. (\ref{eq:phi}), along the NEB path for $12 \times 12 \times 12$ supercells (same color code as in Fig. \ref{fig:Pnma_rot}).
(c) Evolution of polarization components $P_x$ ($=P_y$), and $P_z$ along the NEB path for $12 \times 12 \times 12$ supercells.
Notably, $P_x = P_y$ holds throughout the entire path.
The insets in (c) show the cell-by-cell polarization map of the $Pbnm$, $R3c$, and two representative intermediate states, with arrows representing the polarization component within the $yz$ plane and the background colored squares indicating the octahedral rotations, characterized by $\phi_z$ (same color code as in Fig. \ref{fig:Pnma_rot}).
The cell-by-cell polarization was calculated based on $B$-site-centered 5-atom pseudocubic perovskite unit cells.}
\label{fig:neb}
\end{figure*}

First, we explore the minimum energy path connecting the $Pbnm$ ($a^-a^-c^+$) and $R3c$ ($a^-a^-a^-$) phases using the NEB method.
Although these two phases exhibit different octahedral rotation patterns, they share a common rotation component, $a^-a^-c^0$. 
To construct the initial path, we first aligned the atomic displacements contained in the common rotation component and then performed linear interpolation.
Since the calculations based on the second-principles model are computationally much cheaper than first-principles calculations, it is feasible to perform NEB calculations on large supercells using this model.
Benefiting from this efficiency, we repeated the NEB calculations on supercells of different sizes.

The calculation on $2 \times 2 \times 2$ supercells (with respect to the five-atom perovskite unit cell) converged to a path that exhibits one energy barrier [Fig. \ref{fig:neb}(a)].
Along this path, from $Pbnm$ to $R3c$, the octahedra rotations around the pseudocubic $[001]$ direction (denoted as ``$c$-rotations'') gradually change from in-phase to antiphase, i.e., the rotation direction of every second layer of oxygen octahedra is flipped.
The energy barrier of the NEB path is 40.3 meV/f.u. from the $Pbnm$ side and 14.0 meV/f.u. from the $R3c$ side.

On supercells larger than $2 \times 2 \times 2$, an interesting observation is that the calculations converge to NEB paths with multiple energy barriers. 
Specifically, two energy barriers are observed on the NEB path calculated using $4 \times 4 \times 4$ supercells, three barriers for 6×6×6 supercells, and so on, with $N$ barriers expected for $2N \times 2N \times 2N$ supercells [Fig. \ref{fig:neb}(a)]. 
A close inspection of the structural changes along the NEB paths reveals that each of the energy barriers corresponds to the flipping of one layer of octahedra rotations in the simulation supercell.
In the $2 \times 2 \times 2$ supercell, there is only one layer of octahedra rotation flipped during the change of $c$-rotations from in-phase to antiphase, resulting in one single energy barrier; whereas in larger supercells, more layers of octahedra are allowed to flip independently, and therefore the calculations result in multiple energy barriers.
The multiple energy barriers in larger supercells indicate that along the minimum energy path from $Pbnm$ to $R3c$, different layers of octahedra are flipped one-by-one.

Another interesting observation from the NEB calculations is the decreasing trend of the energy barrier with the increase of the supercell size [Fig. \ref{fig:neb}(a)], which seems to imply a paradox: as the supercell size becomes infinitely large, the barrier for the transition from $R3c$ to $Pbnm$ will tend to be infinitely small.
This could lead to the suspicion that the $R3c$ phase might not be able to remain stable on a macroscopic scale, as even a small perturbation could destabilize $R3c$ in favor of $Pbnm$, which seems to contradict the fact that the $R3c$ phase is dynamically stable and does not show lattice instability at 0 K.
To resolve this paradox, we must consider the localized nature of the layer-by-layer flipping of octahedral rotations along the NEB path, and change our way of computing the energy.
It should be noted that each time one layer of octahedra rotations flips, the energy cost is concentrated locally on that flipping layer, rather than being distributed throughout the entire system. 
From this view, instead of normalizing the energy over the size of the entire system (energy per formula unit), a more appropriate way to represent the energy is to normalize with respect to the cross-sectional area perpendicular to the octahedral rotation axis.

In Fig. \ref{fig:neb}(b), the energy normalized with respect to the cross-sectional area is plotted against the number of flipped layers of the octahedra rotations.
It can be seen that the energy barrier for flipping one layer is approximately the same in the calculations performed on supercells of different sizes.
In this representation, the energy barrier is not expected to decrease toward zero when extrapolating the size of the system to macroscopic scale.
Therefore, our calculations reveal that the minimum energy path connecting the $Pbnm$ and $R3c$ phases consists of consecutive flips of the $c$-rotations, and emphasize the localized nature of the layer-by-layer flips along the path.

Interestingly, going from the nonpolar $Pbnm$ phase to polar $R3c$ phase, the layer-by-layer flipping of oxygen octahedra rotation is also accompanied with a gradual development of the polarization.
Fig. \ref{fig:neb}(c) shows the evolution of the polarization components $P_x$, $P_y$, and $P_z$ along the NEB path, all of which increase as the system goes from $Pbnm$ to $R3c$.
A closer examination of the cell-by-cell polarization [insets in Fig. \ref{fig:neb}(c)] reveals that the development of polarization is spatially non-uniform.
In the intermediate states along the NEB path, in-phase and antiphase $c$-rotations coexist.
In the region with antiphase $c$-rotations, the cell-by-cell polarization vectors are nearly aligned, locally resembling the $R3c$ structure.
By contrast, the region with in-phase $c$-rotations is similar to the $Pbnm$ phase, but exhibits a finite $P_z$, which arises from the requirement to maintain polarization continuity with the adjacent $R3c$-like region.
The presence of a non-zero $P_z$ component in this region accounts for the fact that the net polarization $P_z$ exceeds both $P_x$ and $P_y$.
According to the previous discussion, the transition from $Pbnm$ to $R3c$ along the NEB path can also be interpreted as the movement of the interface between these two phases.
This interfacial motion is accomplished by the local flipping of the oxygen octahedral rotations and the reorientation of the polarization in the unit cells close to the interface.

\subsection{Thermal stability of the $R3c$ phase}

\begin{figure*}[b]
\centering
\includegraphics[scale=0.6]{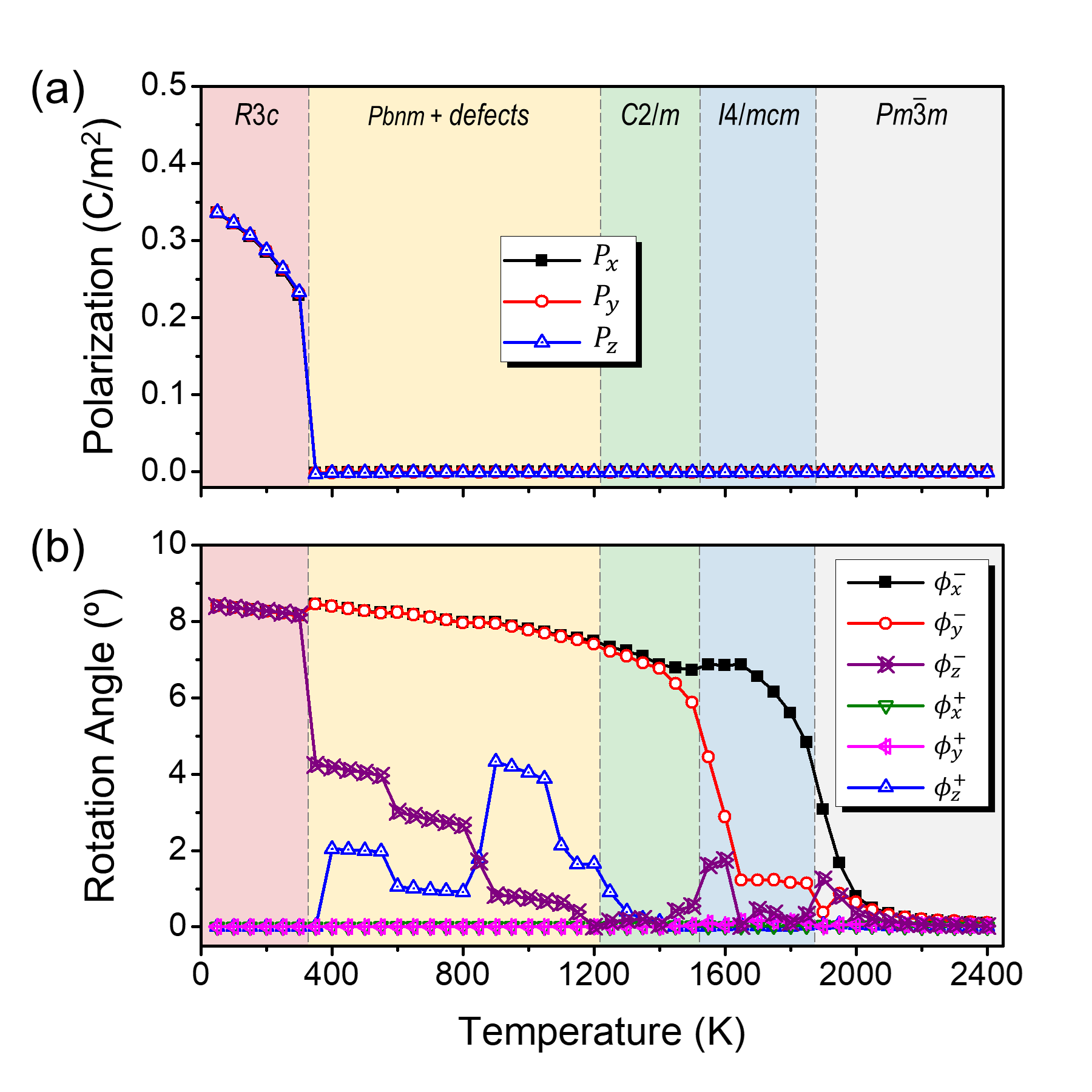}
\caption{Structural phase transitions during the heating process from the $R3c$ phase. 
(a) Polarization, and (b) oxygen octahedral rotations as functions of temperature.
The simulation was performed in $16 \times 16 \times 16$ supercell using the second-principles model of CTO.}
\label{fig:R3c_heating}
\end{figure*}

\begin{figure*}[t]
\centering
\includegraphics[scale=0.50]{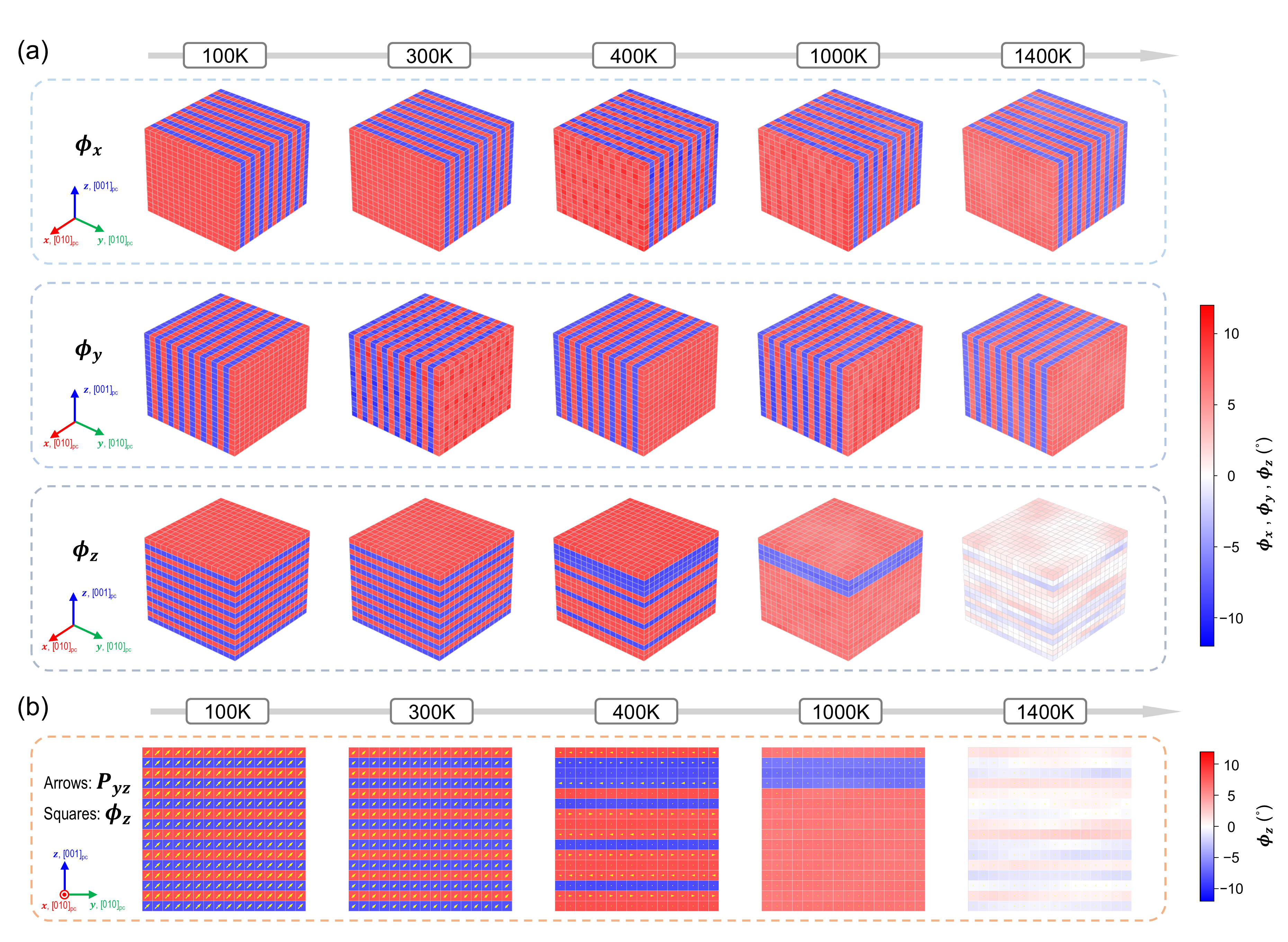}
\caption{Evolution of (a) octahedral rotations and (b) cell-by-cell polarization during the heating process from the $R3c$ phase.
The intensity of the colors in (a) represents the amplitude of $\phi_x$, $\phi_y$, and $\phi_z$ defined by Eq. (\ref{eq:phi}), with red and blue indicating opposite signs.
The arrows in (b) denote the in-plane component of the cell-by-cell polarization component within the $yz$ plane. 
As $P_x = P_y$ holds at all temperatures, the out-of-plane component $P_x$ is not shown.
The background colored squares indicate the octahedral rotations, characterized by $\phi_z$.
The cell-by-cell polarization was calculated based on $B$-site-centered 5-atom pseudocubic perovskite unit cells.}
\label{fig:R3c_rot}
\end{figure*}

Then, assuming it has been induced in some way, we investigate the thermal stability of the bulk $R3c$ phase using the second-principles CTO model by performing HMC heating simulations in $16 \times 16 \times 16$ supercell.
Our model-based calculations confirm that the $R3c$ phase, once stabilized, is remaining stable below a certain finite temperature.
Our results show that the $R3c$ phase, characterized by $P_x = P_y = P_z \neq 0$ and $\phi_x^- = \phi_y^- = \phi_z^- \neq 0$, remains stable up to around 300 K (Fig. \ref{fig:R3c_heating}), consistent with the experimental stabilization of the $R3c$ phase at room temperature \cite{RN461}.
Between 300 K and 350 K, a transition from the $R3c$ phase to some nonpolar structure occurs, as evidenced by the fact that the polarization abruptly drops to zero [Fig. \ref{fig:R3c_heating}(a)].
Above 350 K, the octahedral rotations around two of the pseudocubic axes remain antiphase ($\phi_x^- = \phi_y^- \neq 0$, and $\phi_x^+ = \phi_y^+ = 0$), while the rotations around the third pseudocubic axis change to a complex pattern ($\phi_z^- \neq 0$ and $\phi_z^+ \neq 0$) \cite{RN577}.
As the temperature further increases, both $\phi_z^-$ and $\phi_z^+$ exhibit several jumps [Fig. \ref{fig:R3c_heating}(b)].
When the temperature reaches around 1200 K, the system goes into the $C/2$ phase, and both $\phi_z^-$ and $\phi_z^+$ become zero.
At higher temperatures, the phase transitions are the same as those observed when heating from the $Pnma$ phase (Fig. \ref{fig:Pnma_heating}).

The evolution of octahedral rotations and cell-by-cell polarization with temperature is shown in Fig. \ref{fig:R3c_rot}.
It can be seen that the antiphase rotations around the three pseudocubic axes and the uniform polarization are well preserved up to 300 K.
Above this temperature, the octahedral rotations around the pseudocubic $[001]$ axis, which are characterized by $\phi_z$, are neither in-phase nor antiphase, but exhibit some complex patterns.
These complex patterns can also be regarded as a defective in-phase pattern, as the planes across which $\phi_z$ changes its sign can be viewed as defect planes within a matrix with in-phase rotations.
Accordingly, the polarization mostly disappears with small residual polarization remaining close to the defect planes.
As the temperature further increases, these defect planes are progressively eliminated, the proportion of in-phase rotations increases, and the polarization disappears.
This process occurs through local switching of octahedral rotations, very similar to what we observed in the NEB calculations (Fig. \ref{fig:neb}).
In fact, the jumps in $\phi_z^-$ and $\phi_z^+$ correspond to the local switching of octahedral rotations [Fig. \ref{fig:R3c_heating}(b)].
Therefore, our results reveal that the thermal destabilization of the $R3c$ phase toward the $Pbnm$ phase does not occur as a direct and global transition, but rather happens progressively in a temperature range through consecutive and localized flipping of octahedral rotations.

\subsection{Electric-field induced switching behaviors}

\begin{figure*}[b]
\centering
\includegraphics[scale=0.55]{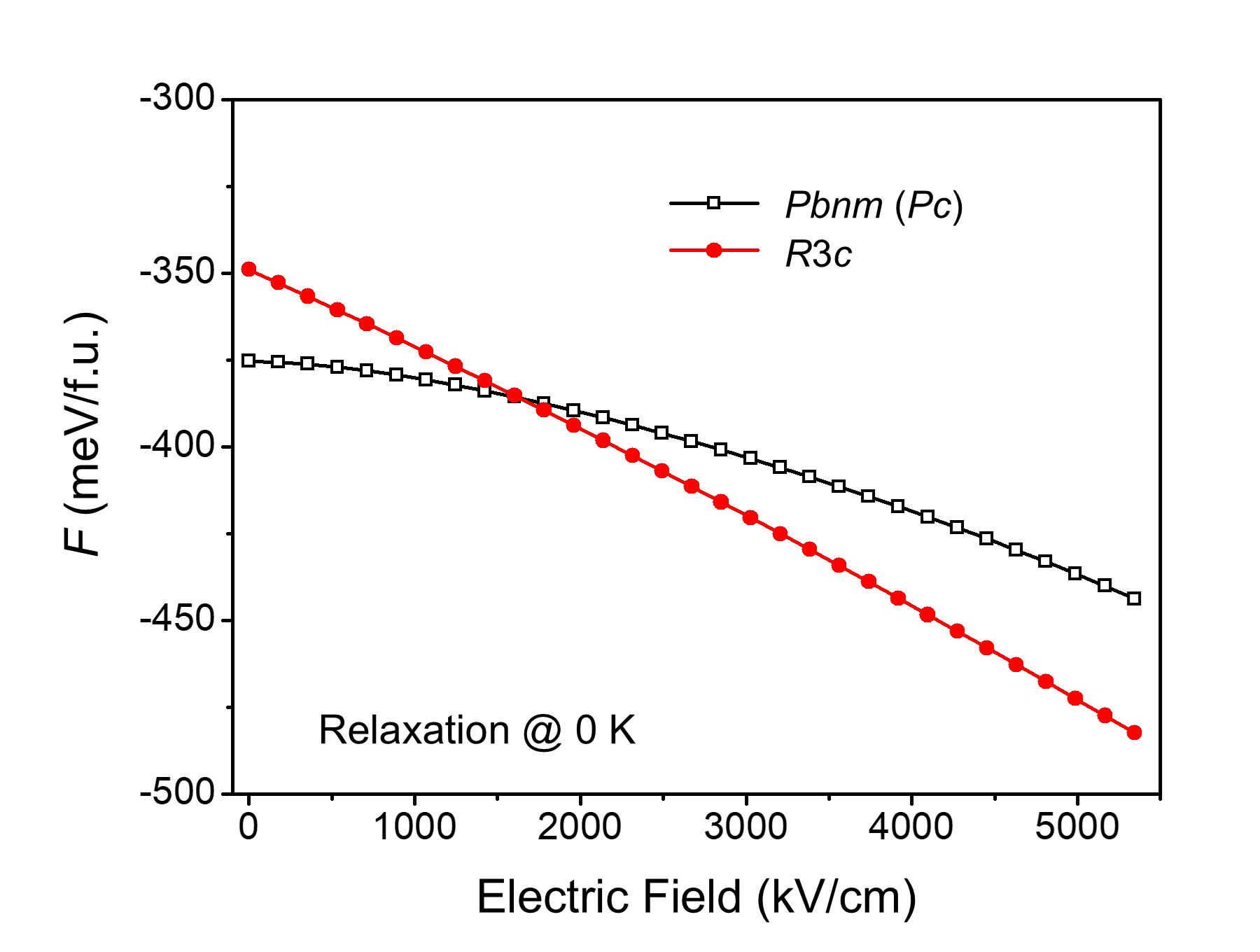}
\caption{
Field-dependent energy $F$ (see text for its definition) of the $Pbnm$ and $R3c$ phases as a function of electric field, calculated from structure relaxations with the electric field applied along pseudocubic $[111]$ direction. 
Note that the symmetry of the $Pbnm$ phase is lowered to $Pc$ under the applied electric field.
}
\label{fig:efield_relax}
\end{figure*}

\begin{figure*}[t]
\centering
\includegraphics[scale=0.50]{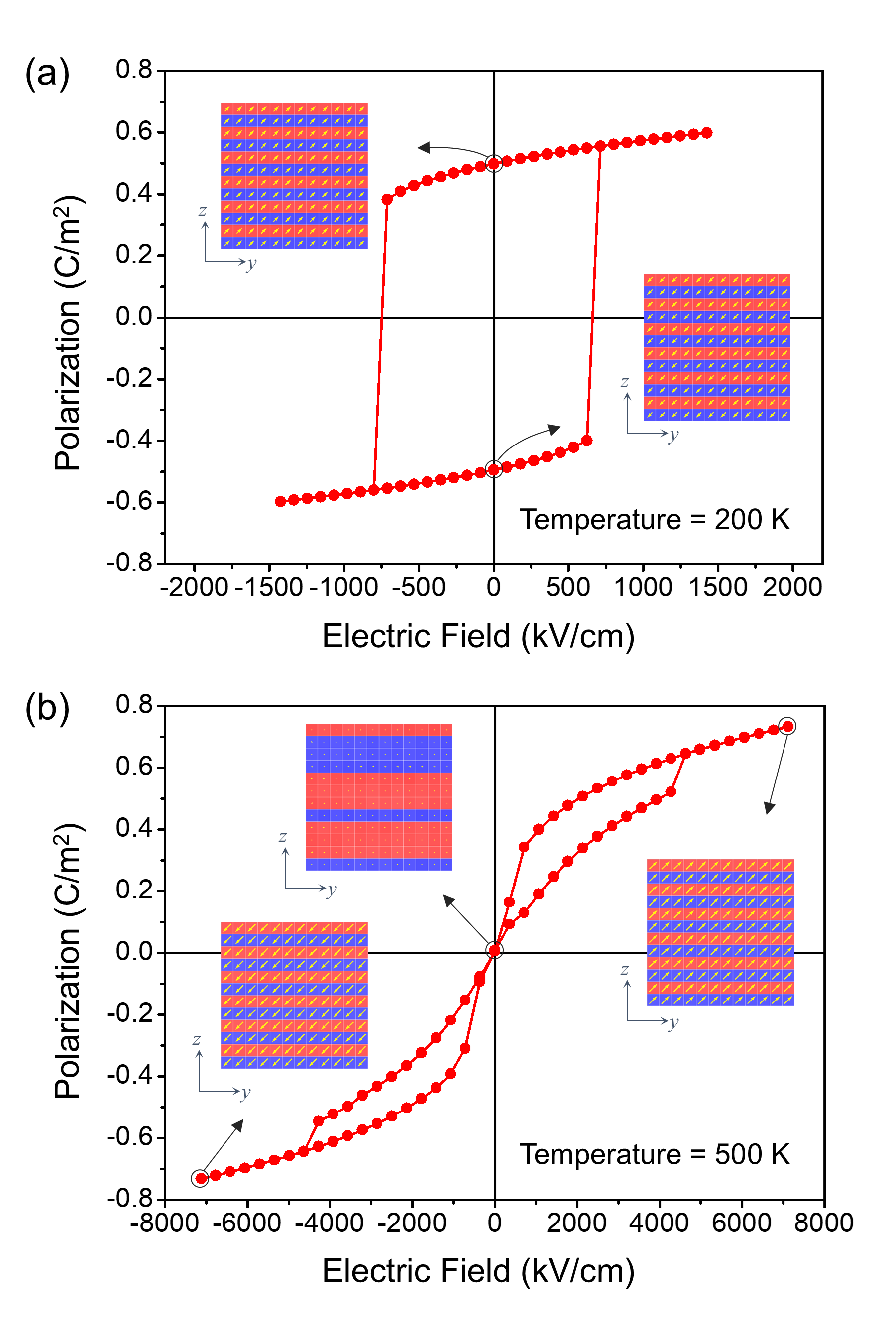}
\caption{
Simulated $P$-$E$ hysteresis loops of CTO at (a) 200 K and (b) 500 K. 
The insets show the cell-by-cell polarization (arrows, representing $P_{yz}$) and octahedral rotations (colored squares, representing $\phi_z$, using the same color code as in Fig. \ref{fig:R3c_rot}) of representative states during the electric field cycling.
The electric field was applied along the pseudocubic $[111]$ direction, and the polarization was projected to the same direction.
The simulations were performed in $12 \times 12 \times 12$ supercells using the second-principles model of CTO.
}
\label{fig:peloop}
\end{figure*}

\begin{figure*}[b]
\centering
\includegraphics[scale=0.50]{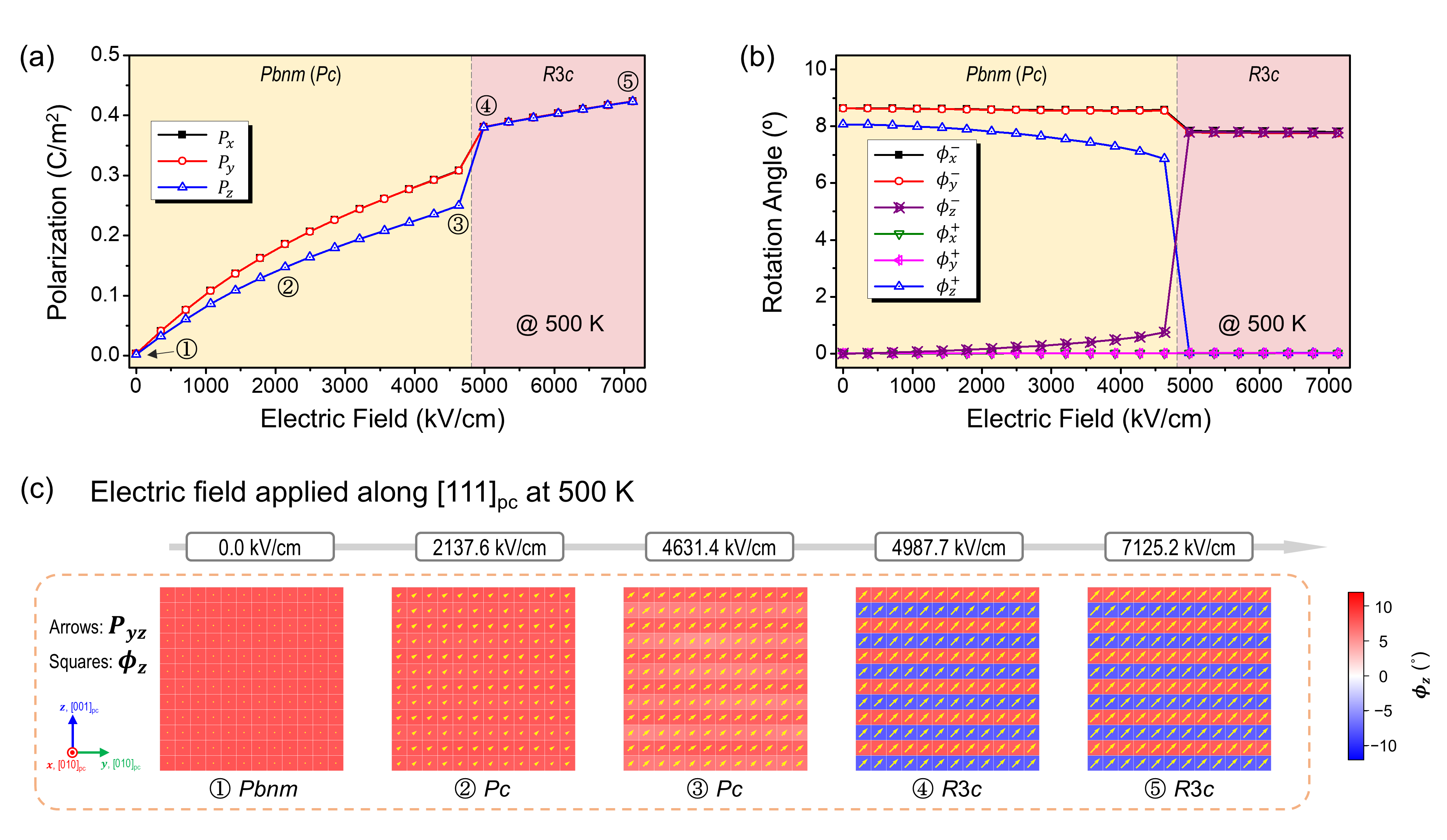}
\caption{
Electric-field induced $Pbnm$-to-$R3c$ phase transition at 500 K with the electric field applied along the pseudocubic $[111]$ direction.
(a) Polarization and (b) octahedral rotation angle as functions of the applied electric field.
(c) Cell-by-cell polarization (arrows, representing $P_{yz}$) and octahedral rotations (colored squares, representing $\phi_z$) during the transition.
Note that the symmetry of the $Pbnm$ phase is lowered to $Pc$ under the applied electric field. 
The simulation was performed in $12 \times 12 \times 12$ supercells using the second-principles model of CTO.
}
\label{fig:efield_500K}
\end{figure*}

The remaining question is to which extent the hidden $R3c$ phase could be eventually stabilized in bulk CTO.
Given the reasonably small energy difference between the $Pbnm$ and $R3c$ phases (see Fig. \ref{fig:neb}) \footnote{We notice that, compared to DFT, the MULTIBINIT model tends to overstabilize the $R3c$ phase respect to the $Pbnm$ phase (see Table I), so the $Pbnm$-to-$R3c$ transition fields reported in this section should be seen as a lower bound.} and the large polarization of the latter, it is reasonable to expect that the $R3c$ phase could be stabilized over the $Pbnm$ phase under appropriate electric field.

To confirm this first at 0 K, Fig. \ref{fig:efield_relax} compares the field-dependent energy functional $F$ defined as $F = U - \Omega (\mathbf{E} \cdot \mathbf{P})$ (where $U$ is the internal energy, $\Omega$ the unit cell volume, $\mathbf{P}$ the macroscopic polarization vector and $\mathbf{E}$ the homogeneous electric field vector), for the $Pbnm$ and $R3c$ phases as a function of electric field \footnote{Strictly speaking, $F$ is not the electric enthalpy (which would include an additional $E^2$ term) but is a proper quantity to minimize to identify the ground state under finite electric field [Phys. Rev. B 63, 155107 (2001)]}.
At zero field, the $Pbnm$ phase exhibits the lowest $F$, but this is changing for increasing fields.
Above 1700 kV/cm, the $R3c$ phase becomes thermodynamically more favorable, suggesting a tendency for a transition from $Pbnm$ to $R3c$ under sufficiently large electric field.
Despite this, the electric field of 1700 kV/cm is not sufficient to induce the $Pbnm$-to-$R3c$ transition.
A larger field is required to overcome the barrier between the two phases (even 5000 kV/cm is not sufficient, the $Pbnm$ phase remains metastable according to Fig. \ref{fig:efield_relax}), and the transition would be very difficult at 0 K because there is no assistance from thermal fluctuations.

To directly simulate the electric-field induced $Pbnm$-to-$R3c$ transition at finite temperature and access polarization-electric field ($P$-$E$) hysteresis loops, we performed HMC simulations while adding the same extra energy term $- \Omega (\mathbf{E} \cdot \mathbf{P})$ as above \cite{RN20}.
This gives rise, in practice, to an extra force on each of the atoms, proportional to the product of the Born effective charge (estimated in the parent cubic phase) and the field. The simulations were performed at two representative temperatures: 200 K, at which the $R3c$ phase is metastable in the absence of electric field, and 500 K, at which the $R3c$ phase becomes dynamically unstable and transforms spontaneously into nonpolar structure in zero field (see Fig. \ref{fig:R3c_heating}).

At 200 K, starting in the metastable $R3c$ phase, our model-based simulations reveal a well-defined single $P$-$E$ loop with a reasonable coercive field of about 750 kV/cm [Fig. \ref{fig:peloop}(a)], confirming the switchable character of the polarization and the ferroelectric nature of the $R3c$ phase. 
This result is consistent with the experimental observation in Ref. \cite{RN461}, where reversible polarization switching was detected by piezoresponse force microscopy (PFM) in thin films. 
This suggests that bulk CTO could potentially be made ferroelectric. However, starting in the $Pbnm$ phase, our calculations reveal that turning the system into the $R3c$ phase would require a field of 3000-5000 kV/cm which is beyond the breakdown field of real samples \cite{CTO_ceramic}. 
Since the model tends to overstabilize the $R3c$ phase, the fact that the transition remains inaccessible even under these conditions indicates that such a field-induced transition is unlikely to occur experimentally.
Alternatively, cooling CTO from 1300 K (in the $C2/m$ phase) under a more reasonable field of 1000 kV/cm might help stabilize the ferroelectric $R3c$ phase at low temperature but would certainly remain very challenging experimentally as well.

At 500 K, the finite-field simulations reveal a double $P$-$E$ hysteresis loop [Fig. \ref{fig:peloop}(b)], fingerprint of an antiferroelectric behavior. 
The vanishing remanent polarization confirms that the polar $R3c$ phase is no longer stable at zero field. 
Instead, the system adopts a nonpolar structure with complex octahedral rotation pattern, which can be regarded as $Pbnm$ structure with planar defects. 
The electric-field induced transition to the $R3c$ phase is first order and identified by an abrupt jump of the polarization and a change in the octahedral rotation pattern from $a^-a^-c^+$ to $a^-a^-a^-$  (Fig. \ref{fig:efield_500K}). 
As such, CTO could be considered as antiferroelectric in this temperature regime. 
However, once again, the electric field for the nonpolar-polar transition (around 4700 kV/cm) clearly exceeds the breakdown field of real samples \cite{CTO_ceramic}, making such behavior hardly accessible experimentally. 
These results are consistent with the fact that, although not a priori impossible, the antiferroelectric-like switching in bulk CTO has never been reported experimentally. All this points out that not only the energetic proximity of the polar and non-polar phases but also the energy barrier between them is key in assessing the antiferroelectric behavior of a compound in a given temperature range.

\section{Conclusions}
In summary, we have reported a second-principles effective interatomic potential for the prototypical perovskite CTO based on Taylor polynomial expansion of the potential energy surface. 
The model is shown to capture a large number of phases, correctly predicting their relative stability, as well as the phonon dispersions for both the nonpolar $Pbnm$ ground state and the hidden ferroelectric $R3c$ phase.
The model-based finite-temperature simulation suggests that upon heating the octahedral rotations around the three pseudocubic axes disappear progressively, with a $C2/m$ ($a^-b^-c^0$) intermediate phase appearing between the $Pnma$ and $I4/mcm$ phases.
Furthermore, the $R3c$ phase is shown to be able to remain stable below a certain finite temperature, and its thermal destabilization process toward the $Pbnm$ ground state is accomplished through consecutive and localized flipping of octahedral rotations, accompanied with a linked local disappearance of the polarization. 
Under electric field, the $Pbnm$-to-$R3c$ phase transition is theoretically feasible within the model, making CTO potentially ferroelectric or antiferroelectric depending on the temperature regime. 
However, the extremely large field required makes it experimentally very challenging to achieve.
Our second-principles model is made freely available and opens perspectives for larger-scale atomistic investigations. It was, for instance, recently used to track and rationalize the appearance of a switching plane in orthorhombic perovskite thin films \cite{Duncan-25}.

\begin{acknowledgments}
This work is supported by the European Union’s Horizon 2020 research and innovation program under grant agreement number 964931 (TSAR) and by F.R.S.-FNRS Belgium under PDR grants T.0107.20 (PROMOSPAN) and T.0128.25 (TOPOTEX). 
H.Z. acknowledges the International Postdoctoral Exchange Fellowship Program (Grant No. PC2020060) of the China Postdoctoral Council, the research IPD-STEMA Program of ULiège (Belgium), and the Fundamental Research Funds for the Central Universities (Grant No. 104972025RSCbs0021) of China.
The authors acknowledge the use of the CECI supercomputer facilities funded by the F.R.S-FNRS (Grant No. 2.5020.1) and of the Tier-1 supercomputer of the Fédération Wallonie-Bruxelles funded by the Walloon Region (Grant No. 1117545).
\end{acknowledgments}

\section*{Data Availability}
The second-principles MULTIBINIT model for CTO is open source and available via the ULiège Open Data Repository \cite{CTO_model}.
Other supporting data are available from the authors upon reasonable request.

\appendix
\section*{Appendix}

In this appendix, Table \ref{tab:coeffs} summarizes all the anharmonic SATs and their coefficient values of the second-principles CTO model developed in this study.

\LTcapwidth=\textwidth
\begin{longtable}{p{1.5cm} p{10cm} p{1.5cm} r}
\label{tab:coeffs} \\
\caption{Symmetry-adapted anharmonic terms of the second-principles effective interatomic potential for $\rm CaTiO_3$. 
The first 60 terms were selected by fitting process, and the rest terms were determined by bounding process.
The unit of the coefficients is in ${\rm hartrees}/{{\rm bohr}^m}$, where $m$ is the order of the atomic displacements in the term. 
Indices in brackets give the position of the cell where the atom is located (if not in the cell $[000]$).} \\
\hline 
No. & Anharmonic term & Order & Value \\ 
\hline 
\endfirsthead
\hline 
No. & Anharmonic term & Order & Value \\ 
\hline 
\endhead
\hline
\endfoot
\hline
\endlastfoot
1 & (Ca$_x$-O1$_x$)(Ca$_x$-O1$_x[\bar{1}00]$)$\eta_1$ & 3 & $-1.3724 \times 10^{-2}$ \\ 
2 & (Ca$_x$-O1$_x$)$^2$(Ca$_y$-O2$_y$)(Ca$_x$-O1$_x[\bar{1}00]$) & 4 & $6.9069 \times 10^{-4}$ \\ 
3 & (Ca$_x$-O1$_x$)$^2$(Ca$_x$-O3$_x[\bar{1}\bar{1}0]$)$^2$ & 4 & $6.2969 \times 10^{-4}$ \\ 
4 & (Ca$_y$-O1$_y$)$^2\eta_2$ & 3 & $-8.6805 \times 10^{-3}$ \\ 
5 & (Ca$_x$-O1$_x$)(Ca$_x$-Ti$_x[00\bar{1}]$)$\eta_1$ & 3 & $6.1934 \times 10^{-3}$ \\ 
6 & (Ca$_x$-O1$_x$)$^2$(Ca$_x$-O3$_x[\bar{1}\bar{1}0]$)(Ca$_x$-Ti$_x[\bar{1}\bar{1}\bar{1}]$) & 4 & $-1.9854 \times 10^{-4}$ \\ 
7 & (Ca$_x$-O1$_x$)(Ca$_z$-Ti$_z[\bar{1}\bar{1}0]$)$\eta_5$ & 3 & $-8.4832 \times 10^{-3}$ \\ 
8 & (Ca$_x$-O1$_x$)(Ca$_x$-O2$_x$)(Ca$_x$-O2$_x[0\bar{1}0]$)(Ca$_x$-O2$_x[0\bar{1}\bar{1}]$) & 4 & $4.8646 \times 10^{-4}$ \\ 
9 & (Ca$_x$-O1$_x$)$^2$(Ca$_z$-O1$_z$)$^2$ & 4 & $2.3068 \times 10^{-3}$ \\ 
10 & (Ca$_x$-O1$_x$)(Ca$_x$-Ti$_x$)(Ca$_x$-O2$_x[0\bar{1}\bar{1}]$)$^2$ & 4 & $-8.4716 \times 10^{-4}$ \\ 
11 & (Ca$_x$-O1$_x$)(Ca$_x$-O2$_x$)(Ca$_x$-Ti$_x[0\bar{1}\bar{1}]$)(Ca$_x$-Ti$_x[\bar{1}\bar{1}\bar{1}]$) & 4 & $1.2559 \times 10^{-3}$ \\ 
12 & (Ca$_x$-O1$_x$)(Ca$_y$-O2$_y$)$\eta_1$ & 3 & $-1.0996 \times 10^{-2}$ \\ 
13 & (Ca$_x$-O1$_x$)(Ca$_y$-O2$_y$)(Ca$_x$-O3$_x$)(Ca$_x$-O1$_x[\bar{1}0\bar{1}]$) & 4 & $-1.0196 \times 10^{-4}$ \\ 
14 & (Ca$_x$-O1$_x$)(Ca$_z$-O1$_z$)(Ca$_z$-Ti$_z[00\bar{1}]$) & 3 & $1.6358 \times 10^{-3}$ \\ 
15 & (Ca$_x$-O1$_x$)(Ca$_x$-O2$_x$)(Ca$_x$-O2$_x[00\bar{1}]$)(Ca$_x$-Ti$_x[\bar{1}\bar{1}0]$) & 4 & $-8.1342 \times 10^{-4}$ \\ 
16 & (Ca$_x$-O1$_x$)(Ca$_y$-O2$_y$)(Ca$_y$-O3$_y$)(Ca$_z$-O1$_z[00\bar{1}]$) & 4 & $-4.9395 \times 10^{-4}$ \\ 
17 & (Ca$_x$-O1$_x$)(Ca$_x$-Ti$_x$)(Ca$_y$-Ti$_y[\bar{1}0\bar{1}]$)$\eta_1$ & 4 & $-1.6225 \times 10^{-2}$ \\ 
18 & (Ca$_x$-O1$_x$)(Ca$_x$-O2$_x$)$^2$(Ca$_y$-O2$_y$) & 4 & $1.4947 \times 10^{-3}$ \\ 
19 & (Ca$_x$-O1$_x$)(Ca$_x$-Ti$_x$)(Ca$_x$-O2$_x$)(Ca$_y$-O2$_y$) & 4 & $-1.1905 \times 10^{-5}$ \\ 
20 & (Ca$_x$-O1$_x$)(Ca$_y$-O2$_y$)(Ca$_y$-Ti$_y$)$^2$ & 4 & $9.2447 \times 10^{-5}$ \\ 
21 & (Ca$_x$-O1$_x$)$^2$(Ca$_x$-O2$_x$)(Ca$_y$-O2$_y$) & 4 & $-1.1206 \times 10^{-3}$ \\ 
22 & (Ca$_x$-O1$_x$)(Ca$_z$-O2$_z$)(Ca$_y$-O3$_y$) & 3 & $-1.5740 \times 10^{-3}$ \\ 
23 & (Ca$_x$-O1$_x$)(Ca$_x$-O1$_x[00\bar{1}]$)$\eta_1$ & 3 & $-1.0746 \times 10^{-3}$ \\ 
24 & (Ca$_x$-O1$_x$)(Ca$_y$-Ti$_y[\bar{1}00]$)(Ca$_x$-Ti$_x[\bar{1}\bar{1}\bar{1}]$)$\eta_3$ & 4 & $-2.7011 \times 10^{-2}$ \\ 
25 & (Ca$_x$-O1$_x$)$^2\eta_1$ & 3 & $-2.5997 \times 10^{-3}$ \\ 
26 & (Ca$_x$-O1$_x$)$^2$(Ca$_z$-O1$_z$)(Ca$_x$-O1$_x[\bar{1}0\bar{1}]$) & 4 & $1.1390 \times 10^{-3}$ \\ 
27 & (Ca$_x$-O1$_x$)(Ca$_y$-O3$_y$)(Ca$_z$-Ti$_z$) & 3 & $9.4888 \times 10^{-4}$ \\ 
28 & (Ca$_y$-O1$_y$)$^2$(Ca$_y$-Ti$_y$) & 3 & $2.8721 \times 10^{-2}$ \\ 
29 & (Ca$_y$-O1$_y$)(Ca$_y$-Ti$_y$)$^2$ & 3 & $-2.8130 \times 10^{-2}$ \\ 
30 & (Ca$_x$-Ti$_x$)$^3$ & 3 & $8.5664 \times 10^{-3}$ \\ 
31 & (Ti$_x$-O1$_x$)$^2$(Ti$_y$-Ti$_y[010]$) & 3 & $3.7061 \times 10^{-3}$ \\ 
32 & (Ti$_x$-O1$_x$)(Ti$_x$-O2$_x$)$^2$(Ti$_x$-O2$_x[100]$) & 4 & $6.9107 \times 10^{-3}$ \\ 
33 & (Ti$_y$-O1$_y$)$^2\eta_2$ & 3 & $-4.0475 \times 10^{-1}$ \\ 
34 & (Ti$_x$-O1$_x$)(Ti$_y$-O2$_y$)(Ti$_x$-O2$_x[100]$)(Ti$_y$-O1$_y[010]$) & 4 & $1.2685 \times 10^{-2}$ \\ 
35 & (Ti$_x$-O1$_x$)(Ti$_x$-Ti$_x[010]$)$\eta_1$ & 3 & $1.6020 \times 10^{-2}$ \\ 
36 & (Ti$_x$-O1$_x$)(Ti$_y$-O1$_y$)(Ti$_x$-Ti$_x[100]$) & 3 & $-8.5879 \times 10^{-4}$ \\ 
37 & (Ti$_x$-O1$_x$)$^2$(Ti$_y$-O1$_y$)$^2$ & 4 & $-1.2741 \times 10^{-2}$ \\ 
38 & (Ti$_x$-O1$_x$)$^3$(Ti$_y$-O2$_y$) & 4 & $-9.8965 \times 10^{-4}$ \\ 
39 & (Ti$_x$-O1$_x$)$^2$(Ti$_y$-O2$_y$)(Ti$_y$-Ti$_y[010]$) & 4 & $-3.6127 \times 10^{-3}$ \\ 
40 & (Ti$_x$-O1$_x$)$^2$(Ti$_z$-O3$_z$) & 3 & $2.7211 \times 10^{-3}$ \\ 
41 & (Ti$_x$-O1$_x$)$^2$(Ti$_x$-O2$_x$) & 3 & $2.1877 \times 10^{-3}$ \\ 
42 & (Ti$_x$-O1$_x$)$^2$(Ti$_z$-O1$_z$)(Ti$_y$-O3$_y$) & 4 & $-8.6731 \times 10^{-4}$ \\ 
43 & (Ca$_x$-Ti$_x$)$^2$(Ca$_y$-Ti$_y$) & 3 & $1.4559 \times 10^{-3}$ \\ 
44 & (Ca$_x$-Ti$_x$)(Ca$_y$-Ti$_y$)(Ca$_z$-Ti$_z$) & 3 & $5.7795 \times 10^{-3}$ \\ 
45 & (Ti$_x$-O1$_x$)$^2\eta_3$ & 3 & $-7.6516 \times 10^{-4}$ \\ 
46 & (Ti$_y$-O1$_y$)(Ti$_y$-O1$_y[010]$)$\eta_2$ & 3 & $8.3555 \times 10^{-2}$ \\ 
47 & (Ti$_x$-O1$_x$)(Ti$_z$-O1$_z$)$\eta_5$ & 3 & $-1.6447 \times 10^{-3}$ \\ 
48 & (Ti$_y$-O1$_y$)(Ti$_x$-O2$_x$)$\eta_1$ & 3 & $-2.1447 \times 10^{-1}$ \\ 
49 & (Ti$_x$-O1$_x$)(Ti$_x$-O1$_x[010]$)(Ti$_y$-O1$_y[010]$) & 3 & $-1.1022 \times 10^{-2}$ \\ 
50 & (Ti$_x$-O1$_x$)(Ti$_x$-O3$_x$)$\eta_4$ & 3 & $2.5619 \times 10^{-3}$ \\ 
51 & (Ti$_x$-O1$_x$)(Ti$_z$-O2$_z$)(Ti$_z$-O1$_z[010]$)$\eta_1$ & 4 & $7.2379 \times 10^{-3}$ \\ 
52 & (Ti$_x$-O1$_x$)(Ti$_x$-O1$_x[010]$)(Ti$_y$-O1$_y[010]$)$^2$ & 4 & $4.9312 \times 10^{-3}$ \\ 
53 & (Ti$_x$-O1$_x$)$^2$(Ti$_y$-O2$_y$)$\eta_3$ & 4 & $1.0173 \times 10^{-1}$ \\ 
54 & (Ti$_x$-O1$_x$)(Ti$_x$-O2$_x$)(Ti$_y$-O2$_y$) & 3 & $-2.9320 \times 10^{-3}$ \\ 
55 & (Ti$_x$-O1$_x$)(Ti$_x$-O3$_x$)(Ti$_z$-Ti$_z[001]$) & 3 & $2.7299 \times 10^{-3}$ \\ 
56 & (Ti$_x$-O1$_x$)(Ti$_x$-O3$_x$)(Ti$_z$-O3$_z$) & 3 & $-3.6066 \times 10^{-3}$ \\ 
57 & (Ti$_x$-O1$_x$)$^2$(Ti$_y$-O3$_y$) & 3 & $-1.0675 \times 10^{-3}$ \\ 
58 & (Ti$_y$-O1$_y$)$^4$ & 4 & $1.4401 \times 10^{-2}$ \\ 
59 & (Ti$_x$-O1$_x$)$^2$(Ti$_y$-O1$_y$)(Ti$_y$-O3$_y$) & 4 & $2.7245 \times 10^{-3}$ \\ 
60 & (Ti$_x$-O1$_x$)$^2$(Ti$_x$-O2$_x$)(Ti$_x$-O1$_x[010]$) & 4 & $1.1044 \times 10^{-3}$ \\ 
61 & (Ca$_x$-O1$_x$)$^2$(Ca$_x$-O1$_x[\bar{1}00]$)$^2$($\eta_1$)$^2$ & 6 & $1.0971 \times 10^{-2}$ \\ 
62 & (Ca$_x$-O1$_x$)$^4$(Ca$_x$-O1$_x[\bar{1}00]$)$^2$($\eta_1$)$^2$ & 8 & $1.3621 \times 10^{-3}$ \\ 
63 & (Ca$_x$-O1$_x$)$^2$(Ca$_x$-O1$_x[\bar{1}00]$)$^2$($\eta_1$)$^4$ & 8 & $6.2923 \times 10^{-1}$ \\ 
64 & (Ca$_x$-O1$_x$)$^2$(Ca$_y$-O2$_y$)$^2$(Ca$_x$-O1$_x[\bar{1}00]$)$^2$ & 6 & $8.0059 \times 10^{-6}$ \\ 
65 & (Ca$_x$-O1$_x$)$^4$(Ca$_y$-O2$_y$)$^2$(Ca$_x$-O1$_x[\bar{1}00]$)$^2$ & 8 & $2.3887 \times 10^{-6}$ \\ 
66 & (Ca$_x$-O1$_x$)$^2$(Ca$_y$-O2$_y$)$^4$(Ca$_x$-O1$_x[\bar{1}00]$)$^2$ & 8 & $3.5855 \times 10^{-6}$ \\ 
67 & (Ca$_y$-O1$_y$)$^4$($\eta_2$)$^2$ & 6 & $4.9040 \times 10^{-3}$ \\ 
68 & (Ca$_y$-O1$_y$)$^2$($\eta_2$)$^4$ & 6 & $9.6879 \times 10^{-1}$ \\ 
69 & (Ca$_y$-O1$_y$)$^6$($\eta_2$)$^2$ & 8 & $3.5439 \times 10^{-3}$ \\ 
70 & (Ca$_y$-O1$_y$)$^2$($\eta_2$)$^6$ & 8 & $1.5768 \times 10^{+2}$ \\ 
71 & (Ca$_x$-O1$_x$)$^2$(Ca$_x$-Ti$_x[00\bar{1}]$)$^2$($\eta_1$)$^2$ & 6 & $3.8419 \times 10^{-3}$ \\ 
72 & (Ca$_x$-O1$_x$)$^4$(Ca$_x$-Ti$_x[00\bar{1}]$)$^2$($\eta_1$)$^2$ & 8 & $1.4068 \times 10^{-3}$ \\ 
73 & (Ca$_x$-O1$_x$)$^2$(Ca$_x$-Ti$_x[00\bar{1}]$)$^4$($\eta_1$)$^2$ & 8 & $2.9621 \times 10^{-3}$ \\ 
74 & (Ca$_x$-O1$_x$)$^2$(Ca$_x$-Ti$_x[00\bar{1}]$)$^2$($\eta_1$)$^4$ & 8 & $5.9197 \times 10^{-1}$ \\ 
75 & (Ca$_x$-O1$_x$)$^2$(Ca$_x$-O3$_x[\bar{1}\bar{1}0]$)$^2$(Ca$_x$-Ti$_x[\bar{1}\bar{1}\bar{1}]$)$^2$ & 6 & $3.3736 \times 10^{-6}$ \\ 
76 & (Ca$_x$-O1$_x$)$^4$(Ca$_x$-O3$_x[\bar{1}\bar{1}0]$)$^2$(Ca$_x$-Ti$_x[\bar{1}\bar{1}\bar{1}]$)$^2$ & 8 & $7.7059 \times 10^{-7}$ \\ 
77 & (Ca$_x$-O1$_x$)$^2$(Ca$_x$-O3$_x[\bar{1}\bar{1}0]$)$^4$(Ca$_x$-Ti$_x[\bar{1}\bar{1}\bar{1}]$)$^2$ & 8 & $6.2836 \times 10^{-7}$ \\ 
78 & (Ca$_x$-O1$_x$)$^2$(Ca$_x$-O3$_x[\bar{1}\bar{1}0]$)$^2$(Ca$_x$-Ti$_x[\bar{1}\bar{1}\bar{1}]$)$^4$ & 8 & $1.1866 \times 10^{-6}$ \\ 
79 & (Ca$_x$-O1$_x$)$^2$(Ca$_z$-Ti$_z[\bar{1}\bar{1}0]$)$^2$($\eta_5$)$^2$ & 6 & $8.8035 \times 10^{-4}$ \\ 
80 & (Ca$_x$-O1$_x$)$^4$(Ca$_z$-Ti$_z[\bar{1}\bar{1}0]$)$^2$($\eta_5$)$^2$ & 8 & $4.4264 \times 10^{-2}$ \\ 
81 & (Ca$_x$-O1$_x$)$^2$(Ca$_z$-Ti$_z[\bar{1}\bar{1}0]$)$^4$($\eta_5$)$^2$ & 8 & $9.5412 \times 10^{-2}$ \\ 
82 & (Ca$_x$-O1$_x$)$^2$(Ca$_z$-Ti$_z[\bar{1}\bar{1}0]$)$^2$($\eta_5$)$^4$ & 8 & $1.1961 \times 10^{+1}$ \\ 
83 & (Ca$_x$-O1$_x$)$^2$(Ca$_x$-O2$_x$)$^2$(Ca$_x$-O2$_x[0\bar{1}0]$)$^2$(Ca$_x$-O2$_x[0\bar{1}\bar{1}]$)$^2$ & 8 & $1.7317 \times 10^{-6}$ \\ 
84 & (Ca$_x$-O1$_x$)$^2$(Ca$_x$-Ti$_x$)$^2$(Ca$_x$-O2$_x[0\bar{1}\bar{1}]$)$^2$ & 6 & $5.0294 \times 10^{-6}$ \\ 
85 & (Ca$_x$-O1$_x$)$^4$(Ca$_x$-Ti$_x$)$^2$(Ca$_x$-O2$_x[0\bar{1}\bar{1}]$)$^2$ & 8 & $9.3472 \times 10^{-7}$ \\ 
86 & (Ca$_x$-O1$_x$)$^2$(Ca$_x$-Ti$_x$)$^4$(Ca$_x$-O2$_x[0\bar{1}\bar{1}]$)$^2$ & 8 & $1.9809 \times 10^{-6}$ \\ 
87 & (Ca$_x$-O1$_x$)$^2$(Ca$_x$-Ti$_x$)$^2$(Ca$_x$-O2$_x[0\bar{1}\bar{1}]$)$^4$ & 8 & $2.4293 \times 10^{-6}$ \\ 
88 & (Ca$_x$-O1$_x$)$^2$(Ca$_x$-O2$_x$)$^2$(Ca$_x$-Ti$_x[0\bar{1}\bar{1}]$)$^2$(Ca$_x$-Ti$_x[\bar{1}\bar{1}\bar{1}]$)$^2$ & 8 & $2.6212 \times 10^{-6}$ \\ 
89 & (Ca$_x$-O1$_x$)$^2$(Ca$_y$-O2$_y$)$^2$($\eta_1$)$^2$ & 6 & $2.6009 \times 10^{-2}$ \\ 
90 & (Ca$_x$-O1$_x$)$^4$(Ca$_y$-O2$_y$)$^2$($\eta_1$)$^2$ & 8 & $5.9379 \times 10^{-3}$ \\ 
91 & (Ca$_x$-O1$_x$)$^2$(Ca$_y$-O2$_y$)$^4$($\eta_1$)$^2$ & 8 & $7.3000 \times 10^{-3}$ \\ 
92 & (Ca$_x$-O1$_x$)$^2$(Ca$_y$-O2$_y$)$^2$($\eta_1$)$^4$ & 8 & $3.7503 \times 10^{+0}$ \\ 
93 & (Ca$_x$-O1$_x$)$^2$(Ca$_y$-O2$_y$)$^2$(Ca$_x$-O3$_x$)$^2$(Ca$_x$-O1$_x[\bar{1}0\bar{1}]$)$^2$ & 8 & $1.6998 \times 10^{-6}$ \\ 
94 & (Ca$_x$-O1$_x$)$^2$(Ca$_z$-O1$_z$)$^2$(Ca$_z$-Ti$_z[00\bar{1}]$)$^2$ & 6 & $1.0990 \times 10^{-5}$ \\ 
95 & (Ca$_x$-O1$_x$)$^4$(Ca$_z$-O1$_z$)$^2$(Ca$_z$-Ti$_z[00\bar{1}]$)$^2$ & 8 & $7.5896 \times 10^{-6}$ \\ 
96 & (Ca$_x$-O1$_x$)$^2$(Ca$_z$-O1$_z$)$^4$(Ca$_z$-Ti$_z[00\bar{1}]$)$^2$ & 8 & $2.2734 \times 10^{-6}$ \\ 
97 & (Ca$_x$-O1$_x$)$^2$(Ca$_z$-O1$_z$)$^2$(Ca$_z$-Ti$_z[00\bar{1}]$)$^4$ & 8 & $3.7913 \times 10^{-6}$ \\ 
98 & (Ca$_x$-O1$_x$)$^2$(Ca$_x$-O2$_x$)$^2$(Ca$_x$-O2$_x[00\bar{1}]$)$^2$(Ca$_x$-Ti$_x[\bar{1}\bar{1}0]$)$^2$ & 8 & $1.6791 \times 10^{-6}$ \\ 
99 & (Ca$_x$-O1$_x$)$^2$(Ca$_y$-O2$_y$)$^2$(Ca$_y$-O3$_y$)$^2$(Ca$_z$-O1$_z[00\bar{1}]$)$^2$ & 8 & $2.3819 \times 10^{-6}$ \\ 
100 & (Ca$_x$-O1$_x$)$^2$(Ca$_x$-Ti$_x$)$^2$(Ca$_y$-Ti$_y[\bar{1}0\bar{1}]$)$^2$($\eta_1$)$^2$ & 8 & $1.1965 \times 10^{-2}$ \\ 
101 & (Ca$_x$-O1$_x$)$^2$(Ca$_x$-O2$_x$)$^2$(Ca$_y$-O2$_y$)$^2$ & 6 & $5.6044 \times 10^{-6}$ \\ 
102 & (Ca$_x$-O1$_x$)$^4$(Ca$_x$-O2$_x$)$^2$(Ca$_y$-O2$_y$)$^2$ & 8 & $1.3097 \times 10^{-6}$ \\ 
103 & (Ca$_x$-O1$_x$)$^2$(Ca$_x$-O2$_x$)$^4$(Ca$_y$-O2$_y$)$^2$ & 8 & $3.3477 \times 10^{-6}$ \\ 
104 & (Ca$_x$-O1$_x$)$^2$(Ca$_x$-O2$_x$)$^2$(Ca$_y$-O2$_y$)$^4$ & 8 & $3.3631 \times 10^{-6}$ \\ 
105 & (Ca$_x$-O1$_x$)$^2$(Ca$_x$-Ti$_x$)$^2$(Ca$_x$-O2$_x$)$^2$(Ca$_y$-O2$_y$)$^2$ & 8 & $3.2527 \times 10^{-6}$ \\ 
106 & (Ca$_x$-O1$_x$)$^2$(Ca$_y$-O2$_y$)$^2$(Ca$_y$-Ti$_y$)$^2$ & 6 & $5.5867 \times 10^{-6}$ \\ 
107 & (Ca$_x$-O1$_x$)$^4$(Ca$_y$-O2$_y$)$^2$(Ca$_y$-Ti$_y$)$^2$ & 8 & $3.1446 \times 10^{-6}$ \\ 
108 & (Ca$_x$-O1$_x$)$^2$(Ca$_y$-O2$_y$)$^4$(Ca$_y$-Ti$_y$)$^2$ & 8 & $1.2295 \times 10^{-6}$ \\ 
109 & (Ca$_x$-O1$_x$)$^2$(Ca$_y$-O2$_y$)$^2$(Ca$_y$-Ti$_y$)$^4$ & 8 & $2.3772 \times 10^{-6}$ \\ 
110 & (Ca$_x$-O1$_x$)$^2$(Ca$_z$-O2$_z$)$^2$(Ca$_y$-O3$_y$)$^2$ & 6 & $1.0720 \times 10^{-5}$ \\ 
111 & (Ca$_x$-O1$_x$)$^4$(Ca$_z$-O2$_z$)$^2$(Ca$_y$-O3$_y$)$^2$ & 8 & $2.0367 \times 10^{-6}$ \\ 
112 & (Ca$_x$-O1$_x$)$^2$(Ca$_x$-O1$_x[00\bar{1}]$)$^2$($\eta_1$)$^2$ & 6 & $2.7274 \times 10^{-3}$ \\ 
113 & (Ca$_x$-O1$_x$)$^4$(Ca$_x$-O1$_x[00\bar{1}]$)$^2$($\eta_1$)$^2$ & 8 & $6.4797 \times 10^{-4}$ \\ 
114 & (Ca$_x$-O1$_x$)$^2$(Ca$_x$-O1$_x[00\bar{1}]$)$^2$($\eta_1$)$^4$ & 8 & $3.2113 \times 10^{-1}$ \\ 
115 & (Ca$_x$-O1$_x$)$^2$(Ca$_y$-Ti$_y[\bar{1}00]$)$^2$(Ca$_x$-Ti$_x[\bar{1}\bar{1}\bar{1}]$)$^2$($\eta_3$)$^2$ & 8 & $2.7835 \times 10^{-2}$ \\ 
116 & (Ca$_x$-O1$_x$)$^4$($\eta_1$)$^2$ & 6 & $8.7250 \times 10^{-4}$ \\ 
117 & (Ca$_x$-O1$_x$)$^2$($\eta_1$)$^4$ & 6 & $2.4416 \times 10^{-1}$ \\ 
118 & (Ca$_x$-O1$_x$)$^6$($\eta_1$)$^2$ & 8 & $3.2493 \times 10^{-4}$ \\ 
119 & (Ca$_x$-O1$_x$)$^2$($\eta_1$)$^6$ & 8 & $3.7745 \times 10^{+1}$ \\ 
120 & (Ca$_x$-O1$_x$)$^2$(Ca$_z$-O1$_z$)$^2$(Ca$_x$-O1$_x[\bar{1}0\bar{1}]$)$^2$ & 6 & $9.5747 \times 10^{-6}$ \\ 
121 & (Ca$_x$-O1$_x$)$^4$(Ca$_z$-O1$_z$)$^2$(Ca$_x$-O1$_x[\bar{1}0\bar{1}]$)$^2$ & 8 & $3.7364 \times 10^{-6}$ \\ 
122 & (Ca$_x$-O1$_x$)$^2$(Ca$_z$-O1$_z$)$^4$(Ca$_x$-O1$_x[\bar{1}0\bar{1}]$)$^2$ & 8 & $7.9272 \times 10^{-6}$ \\ 
123 & (Ca$_x$-O1$_x$)$^2$(Ca$_z$-O1$_z$)$^2$(Ca$_x$-O1$_x[\bar{1}0\bar{1}]$)$^4$ & 8 & $4.8418 \times 10^{-6}$ \\ 
124 & (Ca$_x$-O1$_x$)$^2$(Ca$_y$-O3$_y$)$^2$(Ca$_z$-Ti$_z$)$^2$ & 6 & $1.4916 \times 10^{-5}$ \\ 
125 & (Ca$_x$-O1$_x$)$^4$(Ca$_y$-O3$_y$)$^2$(Ca$_z$-Ti$_z$)$^2$ & 8 & $7.9233 \times 10^{-6}$ \\ 
126 & (Ca$_x$-O1$_x$)$^2$(Ca$_y$-O3$_y$)$^4$(Ca$_z$-Ti$_z$)$^2$ & 8 & $6.4708 \times 10^{-6}$ \\ 
127 & (Ca$_x$-O1$_x$)$^2$(Ca$_y$-O3$_y$)$^2$(Ca$_z$-Ti$_z$)$^4$ & 8 & $1.7000 \times 10^{-5}$ \\ 
128 & (Ca$_y$-O1$_y$)$^4$(Ca$_y$-Ti$_y$)$^2$ & 6 & $4.7736 \times 10^{-6}$ \\ 
129 & (Ca$_y$-O1$_y$)$^2$(Ca$_y$-Ti$_y$)$^4$ & 6 & $4.0275 \times 10^{-6}$ \\ 
130 & (Ca$_y$-O1$_y$)$^6$(Ca$_y$-Ti$_y$)$^2$ & 8 & $1.5003 \times 10^{-6}$ \\ 
131 & (Ca$_y$-O1$_y$)$^2$(Ca$_y$-Ti$_y$)$^6$ & 8 & $9.7559 \times 10^{-7}$ \\ 
132 & (Ca$_y$-O1$_y$)$^4$(Ca$_y$-Ti$_y$)$^4$ & 8 & $1.2277 \times 10^{-6}$ \\ 
133 & (Ca$_x$-Ti$_x$)$^6$ & 6 & $2.8511 \times 10^{-6}$ \\ 
134 & (Ca$_x$-Ti$_x$)$^8$ & 8 & $6.7816 \times 10^{-7}$ \\ 
135 & (Ti$_x$-O1$_x$)$^4$(Ti$_y$-Ti$_y[010]$)$^2$ & 6 & $2.3715 \times 10^{-4}$ \\ 
136 & (Ti$_x$-O1$_x$)$^2$(Ti$_y$-Ti$_y[010]$)$^4$ & 6 & $8.6194 \times 10^{-4}$ \\ 
137 & (Ti$_x$-O1$_x$)$^6$(Ti$_y$-Ti$_y[010]$)$^2$ & 8 & $9.2116 \times 10^{-5}$ \\ 
138 & (Ti$_x$-O1$_x$)$^2$(Ti$_y$-Ti$_y[010]$)$^6$ & 8 & $1.3899 \times 10^{-3}$ \\ 
139 & (Ti$_x$-O1$_x$)$^4$(Ti$_y$-Ti$_y[010]$)$^4$ & 8 & $4.0564 \times 10^{-4}$ \\ 
140 & (Ti$_x$-O1$_x$)$^2$(Ti$_x$-O2$_x$)$^2$(Ti$_x$-O2$_x[100]$)$^2$ & 6 & $4.7600 \times 10^{-4}$ \\ 
141 & (Ti$_x$-O1$_x$)$^4$(Ti$_x$-O2$_x$)$^2$(Ti$_x$-O2$_x[100]$)$^2$ & 8 & $3.3572 \times 10^{-4}$ \\ 
142 & (Ti$_x$-O1$_x$)$^2$(Ti$_x$-O2$_x$)$^4$(Ti$_x$-O2$_x[100]$)$^2$ & 8 & $5.8332 \times 10^{-4}$ \\ 
143 & (Ti$_y$-O1$_y$)$^4$($\eta_2$)$^2$ & 6 & $5.2483 \times 10^{-2}$ \\ 
144 & (Ti$_y$-O1$_y$)$^2$($\eta_2$)$^4$ & 6 & $3.2333 \times 10^{+0}$ \\ 
145 & (Ti$_y$-O1$_y$)$^6$($\eta_2$)$^2$ & 8 & $1.6579 \times 10^{-1}$ \\ 
146 & (Ti$_y$-O1$_y$)$^2$($\eta_2$)$^6$ & 8 & $5.8658 \times 10^{+2}$ \\ 
147 & (Ti$_x$-O1$_x$)$^2$(Ti$_y$-O2$_y$)$^2$(Ti$_x$-O2$_x[100]$)$^2$(Ti$_y$-O1$_y[010]$)$^2$ & 8 & $2.3385 \times 10^{-3}$ \\ 
148 & (Ti$_x$-O1$_x$)$^2$(Ti$_x$-Ti$_x[010]$)$^2$($\eta_1$)$^2$ & 6 & $4.4700 \times 10^{-2}$ \\ 
149 & (Ti$_x$-O1$_x$)$^4$(Ti$_x$-Ti$_x[010]$)$^2$($\eta_1$)$^2$ & 8 & $3.2608 \times 10^{-2}$ \\ 
150 & (Ti$_x$-O1$_x$)$^2$(Ti$_x$-Ti$_x[010]$)$^4$($\eta_1$)$^2$ & 8 & $4.6595 \times 10^{-2}$ \\ 
151 & (Ti$_x$-O1$_x$)$^2$(Ti$_x$-Ti$_x[010]$)$^2$($\eta_1$)$^4$ & 8 & $4.7307 \times 10^{+2}$ \\ 
152 & (Ti$_x$-O1$_x$)$^2$(Ti$_y$-O1$_y$)$^2$(Ti$_x$-Ti$_x[100]$)$^2$ & 6 & $1.2930 \times 10^{-3}$ \\ 
153 & (Ti$_x$-O1$_x$)$^4$(Ti$_y$-O1$_y$)$^2$(Ti$_x$-Ti$_x[100]$)$^2$ & 8 & $9.5946 \times 10^{-4}$ \\ 
154 & (Ti$_x$-O1$_x$)$^2$(Ti$_y$-O1$_y$)$^4$(Ti$_x$-Ti$_x[100]$)$^2$ & 8 & $5.4837 \times 10^{-3}$ \\ 
155 & (Ti$_x$-O1$_x$)$^2$(Ti$_y$-O1$_y$)$^2$(Ti$_x$-Ti$_x[100]$)$^4$ & 8 & $3.9716 \times 10^{-3}$ \\ 
156 & (Ti$_x$-O1$_x$)$^4$(Ti$_y$-O1$_y$)$^2$ & 6 & $1.2697 \times 10^{-4}$ \\ 
157 & (Ti$_x$-O1$_x$)$^2$(Ti$_y$-O1$_y$)$^4$ & 6 & $4.9883 \times 10^{-4}$ \\ 
158 & (Ti$_x$-O1$_x$)$^6$(Ti$_y$-O1$_y$)$^2$ & 8 & $7.4888 \times 10^{-5}$ \\ 
159 & (Ti$_x$-O1$_x$)$^2$(Ti$_y$-O1$_y$)$^6$ & 8 & $1.3398 \times 10^{-3}$ \\ 
160 & (Ti$_x$-O1$_x$)$^4$(Ti$_y$-O1$_y$)$^4$ & 8 & $4.3381 \times 10^{-4}$ \\ 
161 & (Ti$_x$-O1$_x$)$^4$(Ti$_y$-O2$_y$)$^2$ & 6 & $5.3952 \times 10^{-6}$ \\ 
162 & (Ti$_x$-O1$_x$)$^6$(Ti$_y$-O2$_y$)$^2$ & 8 & $3.0225 \times 10^{-6}$ \\ 
163 & (Ti$_x$-O1$_x$)$^4$(Ti$_y$-O2$_y$)$^4$ & 8 & $7.3968 \times 10^{-6}$ \\ 
164 & (Ti$_x$-O1$_x$)$^2$(Ti$_y$-O2$_y$)$^2$(Ti$_y$-Ti$_y[010]$)$^2$ & 6 & $9.6803 \times 10^{-5}$ \\ 
165 & (Ti$_x$-O1$_x$)$^4$(Ti$_y$-O2$_y$)$^2$(Ti$_y$-Ti$_y[010]$)$^2$ & 8 & $6.3089 \times 10^{-5}$ \\ 
166 & (Ti$_x$-O1$_x$)$^2$(Ti$_y$-O2$_y$)$^4$(Ti$_y$-Ti$_y[010]$)$^2$ & 8 & $5.8123 \times 10^{-5}$ \\ 
167 & (Ti$_x$-O1$_x$)$^2$(Ti$_y$-O2$_y$)$^2$(Ti$_y$-Ti$_y[010]$)$^4$ & 8 & $2.8154 \times 10^{-4}$ \\ 
168 & (Ti$_x$-O1$_x$)$^4$(Ti$_z$-O3$_z$)$^2$ & 6 & $8.1814 \times 10^{-5}$ \\ 
169 & (Ti$_x$-O1$_x$)$^2$(Ti$_z$-O3$_z$)$^4$ & 6 & $3.1377 \times 10^{-4}$ \\ 
170 & (Ti$_x$-O1$_x$)$^6$(Ti$_z$-O3$_z$)$^2$ & 8 & $5.2636 \times 10^{-5}$ \\ 
171 & (Ti$_x$-O1$_x$)$^2$(Ti$_z$-O3$_z$)$^6$ & 8 & $9.6020 \times 10^{-4}$ \\ 
172 & (Ti$_x$-O1$_x$)$^4$(Ti$_z$-O3$_z$)$^4$ & 8 & $3.2109 \times 10^{-4}$ \\ 
173 & (Ti$_x$-O1$_x$)$^4$(Ti$_x$-O2$_x$)$^2$ & 6 & $3.1036 \times 10^{-5}$ \\ 
174 & (Ti$_x$-O1$_x$)$^2$(Ti$_x$-O2$_x$)$^4$ & 6 & $1.3183 \times 10^{-4}$ \\ 
175 & (Ti$_x$-O1$_x$)$^6$(Ti$_x$-O2$_x$)$^2$ & 8 & $1.5050 \times 10^{-5}$ \\ 
176 & (Ti$_x$-O1$_x$)$^2$(Ti$_x$-O2$_x$)$^6$ & 8 & $3.5092 \times 10^{-4}$ \\ 
177 & (Ti$_x$-O1$_x$)$^4$(Ti$_x$-O2$_x$)$^4$ & 8 & $9.5252 \times 10^{-5}$ \\ 
178 & (Ti$_x$-O1$_x$)$^2$(Ti$_z$-O1$_z$)$^2$(Ti$_y$-O3$_y$)$^2$ & 6 & $1.2373 \times 10^{-5}$ \\ 
179 & (Ti$_x$-O1$_x$)$^4$(Ti$_z$-O1$_z$)$^2$(Ti$_y$-O3$_y$)$^2$ & 8 & $1.4238 \times 10^{-5}$ \\ 
180 & (Ti$_x$-O1$_x$)$^2$(Ti$_z$-O1$_z$)$^4$(Ti$_y$-O3$_y$)$^2$ & 8 & $1.2263 \times 10^{-5}$ \\ 
181 & (Ti$_x$-O1$_x$)$^2$(Ti$_z$-O1$_z$)$^2$(Ti$_y$-O3$_y$)$^4$ & 8 & $1.0752 \times 10^{-5}$ \\ 
182 & (Ca$_x$-Ti$_x$)$^4$(Ca$_y$-Ti$_y$)$^2$ & 6 & $1.5505 \times 10^{-5}$ \\ 
183 & (Ca$_x$-Ti$_x$)$^6$(Ca$_y$-Ti$_y$)$^2$ & 8 & $5.6103 \times 10^{-6}$ \\ 
184 & (Ca$_x$-Ti$_x$)$^4$(Ca$_y$-Ti$_y$)$^4$ & 8 & $3.4553 \times 10^{-5}$ \\ 
185 & (Ca$_x$-Ti$_x$)$^2$(Ca$_y$-Ti$_y$)$^2$(Ca$_z$-Ti$_z$)$^2$ & 6 & $3.4693 \times 10^{-4}$ \\ 
186 & (Ca$_x$-Ti$_x$)$^4$(Ca$_y$-Ti$_y$)$^2$(Ca$_z$-Ti$_z$)$^2$ & 8 & $9.4308 \times 10^{-5}$ \\ 
187 & (Ti$_x$-O1$_x$)$^4$($\eta_3$)$^2$ & 6 & $3.1747 \times 10^{-2}$ \\ 
188 & (Ti$_x$-O1$_x$)$^2$($\eta_3$)$^4$ & 6 & $8.0398 \times 10^{+0}$ \\ 
189 & (Ti$_x$-O1$_x$)$^6$($\eta_3$)$^2$ & 8 & $1.4289 \times 10^{-2}$ \\ 
190 & (Ti$_x$-O1$_x$)$^2$($\eta_3$)$^6$ & 8 & $2.3453 \times 10^{+3}$ \\ 
191 & (Ti$_y$-O1$_y$)$^2$(Ti$_y$-O1$_y[010]$)$^2$($\eta_2$)$^2$ & 6 & $1.0055 \times 10^{-1}$ \\ 
192 & (Ti$_y$-O1$_y$)$^4$(Ti$_y$-O1$_y[010]$)$^2$($\eta_2$)$^2$ & 8 & $1.6107 \times 10^{-1}$ \\ 
193 & (Ti$_y$-O1$_y$)$^2$(Ti$_y$-O1$_y[010]$)$^2$($\eta_2$)$^4$ & 8 & $1.9399 \times 10^{+1}$ \\ 
194 & (Ti$_x$-O1$_x$)$^2$(Ti$_z$-O1$_z$)$^2$($\eta_5$)$^2$ & 6 & $7.5133 \times 10^{-2}$ \\ 
195 & (Ti$_x$-O1$_x$)$^4$(Ti$_z$-O1$_z$)$^2$($\eta_5$)$^2$ & 8 & $5.9098 \times 10^{-2}$ \\ 
196 & (Ti$_x$-O1$_x$)$^2$(Ti$_z$-O1$_z$)$^2$($\eta_5$)$^4$ & 8 & $2.6607 \times 10^{+1}$ \\ 
197 & (Ti$_y$-O1$_y$)$^2$(Ti$_x$-O2$_x$)$^2$($\eta_1$)$^2$ & 6 & $4.2302 \times 10^{-2}$ \\ 
198 & (Ti$_y$-O1$_y$)$^4$(Ti$_x$-O2$_x$)$^2$($\eta_1$)$^2$ & 8 & $2.1685 \times 10^{-1}$ \\ 
199 & (Ti$_y$-O1$_y$)$^2$(Ti$_x$-O2$_x$)$^4$($\eta_1$)$^2$ & 8 & $1.6339 \times 10^{-1}$ \\ 
200 & (Ti$_y$-O1$_y$)$^2$(Ti$_x$-O2$_x$)$^2$($\eta_1$)$^4$ & 8 & $1.5507 \times 10^{+1}$ \\ 
201 & (Ti$_x$-O1$_x$)$^2$(Ti$_x$-O1$_x[010]$)$^2$(Ti$_y$-O1$_y[010]$)$^2$ & 6 & $2.8626 \times 10^{-4}$ \\ 
202 & (Ti$_x$-O1$_x$)$^4$(Ti$_x$-O1$_x[010]$)$^2$(Ti$_y$-O1$_y[010]$)$^2$ & 8 & $3.1933 \times 10^{-4}$ \\ 
203 & (Ti$_x$-O1$_x$)$^2$(Ti$_x$-O1$_x[010]$)$^4$(Ti$_y$-O1$_y[010]$)$^2$ & 8 & $2.9348 \times 10^{-4}$ \\ 
204 & (Ti$_x$-O1$_x$)$^2$(Ti$_x$-O1$_x[010]$)$^2$(Ti$_y$-O1$_y[010]$)$^4$ & 8 & $9.7729 \times 10^{-4}$ \\ 
205 & (Ti$_x$-O1$_x$)$^2$(Ti$_x$-O3$_x$)$^2$($\eta_4$)$^2$ & 6 & $2.2688 \times 10^{-2}$ \\ 
206 & (Ti$_x$-O1$_x$)$^4$(Ti$_x$-O3$_x$)$^2$($\eta_4$)$^2$ & 8 & $1.8052 \times 10^{-2}$ \\ 
207 & (Ti$_x$-O1$_x$)$^2$(Ti$_x$-O3$_x$)$^2$($\eta_4$)$^4$ & 8 & $6.2946 \times 10^{+0}$ \\ 
208 & (Ti$_x$-O1$_x$)$^2$(Ti$_z$-O2$_z$)$^2$(Ti$_z$-O1$_z[010]$)$^2$($\eta_1$)$^2$ & 8 & $8.2202 \times 10^{-2}$ \\ 
209 & (Ti$_x$-O1$_x$)$^2$(Ti$_y$-O2$_y$)$^2$($\eta_3$)$^2$ & 6 & $4.5744 \times 10^{-2}$ \\ 
210 & (Ti$_x$-O1$_x$)$^4$(Ti$_y$-O2$_y$)$^2$($\eta_3$)$^2$ & 8 & $1.3179 \times 10^{-2}$ \\ 
211 & (Ti$_x$-O1$_x$)$^2$(Ti$_y$-O2$_y$)$^2$($\eta_3$)$^4$ & 8 & $2.9093 \times 10^{+1}$ \\ 
212 & (Ti$_x$-O1$_x$)$^2$(Ti$_x$-O2$_x$)$^2$(Ti$_y$-O2$_y$)$^2$ & 6 & $6.4305 \times 10^{-5}$ \\ 
213 & (Ti$_x$-O1$_x$)$^4$(Ti$_x$-O2$_x$)$^2$(Ti$_y$-O2$_y$)$^2$ & 8 & $3.8170 \times 10^{-5}$ \\ 
214 & (Ti$_x$-O1$_x$)$^2$(Ti$_x$-O2$_x$)$^4$(Ti$_y$-O2$_y$)$^2$ & 8 & $2.2516 \times 10^{-4}$ \\ 
215 & (Ti$_x$-O1$_x$)$^2$(Ti$_x$-O2$_x$)$^2$(Ti$_y$-O2$_y$)$^4$ & 8 & $5.0296 \times 10^{-5}$ \\ 
216 & (Ti$_x$-O1$_x$)$^2$(Ti$_x$-O3$_x$)$^2$(Ti$_z$-Ti$_z[001]$)$^2$ & 6 & $1.7971 \times 10^{-4}$ \\ 
217 & (Ti$_x$-O1$_x$)$^4$(Ti$_x$-O3$_x$)$^2$(Ti$_z$-Ti$_z[001]$)$^2$ & 8 & $1.7497 \times 10^{-4}$ \\ 
218 & (Ti$_x$-O1$_x$)$^2$(Ti$_x$-O3$_x$)$^4$(Ti$_z$-Ti$_z[001]$)$^2$ & 8 & $1.1679 \times 10^{-4}$ \\ 
219 & (Ti$_x$-O1$_x$)$^2$(Ti$_x$-O3$_x$)$^2$(Ti$_z$-Ti$_z[001]$)$^4$ & 8 & $6.3681 \times 10^{-4}$ \\ 
220 & (Ti$_x$-O1$_x$)$^2$(Ti$_x$-O3$_x$)$^2$(Ti$_z$-O3$_z$)$^2$ & 6 & $1.0818 \times 10^{-4}$ \\ 
221 & (Ti$_x$-O1$_x$)$^4$(Ti$_x$-O3$_x$)$^2$(Ti$_z$-O3$_z$)$^2$ & 8 & $1.0324 \times 10^{-4}$ \\ 
222 & (Ti$_x$-O1$_x$)$^2$(Ti$_x$-O3$_x$)$^4$(Ti$_z$-O3$_z$)$^2$ & 8 & $9.1903 \times 10^{-5}$ \\ 
223 & (Ti$_x$-O1$_x$)$^2$(Ti$_x$-O3$_x$)$^2$(Ti$_z$-O3$_z$)$^4$ & 8 & $4.3498 \times 10^{-4}$ \\ 
224 & (Ti$_x$-O1$_x$)$^4$(Ti$_y$-O3$_y$)$^2$ & 6 & $8.0599 \times 10^{-6}$ \\ 
225 & (Ti$_x$-O1$_x$)$^2$(Ti$_y$-O3$_y$)$^4$ & 6 & $7.6696 \times 10^{-6}$ \\ 
226 & (Ti$_x$-O1$_x$)$^6$(Ti$_y$-O3$_y$)$^2$ & 8 & $5.9718 \times 10^{-6}$ \\ 
227 & (Ti$_x$-O1$_x$)$^2$(Ti$_y$-O3$_y$)$^6$ & 8 & $5.2529 \times 10^{-6}$ \\ 
228 & (Ti$_x$-O1$_x$)$^4$(Ti$_y$-O3$_y$)$^4$ & 8 & $7.6262 \times 10^{-6}$ \\ 
229 & (Ti$_x$-O1$_x$)$^2$(Ti$_y$-O1$_y$)$^2$(Ti$_y$-O3$_y$)$^2$ & 6 & $7.5187 \times 10^{-5}$ \\ 
230 & (Ti$_x$-O1$_x$)$^4$(Ti$_y$-O1$_y$)$^2$(Ti$_y$-O3$_y$)$^2$ & 8 & $6.9023 \times 10^{-5}$ \\ 
231 & (Ti$_x$-O1$_x$)$^2$(Ti$_y$-O1$_y$)$^4$(Ti$_y$-O3$_y$)$^2$ & 8 & $2.6373 \times 10^{-4}$ \\ 
232 & (Ti$_x$-O1$_x$)$^2$(Ti$_y$-O1$_y$)$^2$(Ti$_y$-O3$_y$)$^4$ & 8 & $5.6278 \times 10^{-5}$ \\ 
233 & (Ti$_x$-O1$_x$)$^2$(Ti$_x$-O2$_x$)$^2$(Ti$_x$-O1$_x[010]$)$^2$ & 6 & $1.9509 \times 10^{-4}$ \\ 
234 & (Ti$_x$-O1$_x$)$^4$(Ti$_x$-O2$_x$)$^2$(Ti$_x$-O1$_x[010]$)$^2$ & 8 & $1.0794 \times 10^{-4}$ \\ 
235 & (Ti$_x$-O1$_x$)$^2$(Ti$_x$-O2$_x$)$^4$(Ti$_x$-O1$_x[010]$)$^2$ & 8 & $5.9611 \times 10^{-4}$ \\ 
236 & ($\eta_1$)$^6$ & 6 & $1.0000 \times 10^{+0}$ \\ 
237 & ($\eta_1$)$^8$ & 8 & $1.0000 \times 10^{+0}$ \\ 
238 & ($\eta_4$)$^6$ & 6 & $1.0000 \times 10^{+0}$ \\ 
239 & ($\eta_4$)$^8$ & 8 & $1.0000 \times 10^{+0}$ \\ 
\end{longtable}

\bibliography{references}
\bibliographystyle{unsrt}

\end{document}